# A Systematic Literature Review on Safety of the Intended Functionality for Automated Driving Systems

**Milin Patel and Rolf Jung** Institute for Driver Assistance and Connected Mobility

**Marzana Khatun** Robert Bosch Engineering

## Abstract

In the automobile industry, ensuring the safety of automated vehicles equipped with the automated driving system (ADS) is becoming a significant focus due to the increasing development and deployment of automated driving. Automated driving depends on sensing both the external and internal environments of a vehicle, utilizing perception sensors and algorithms, and electrical/electronic (E/E) systems for situational awareness and response. ISO 21448 is the standard for Safety of the Intended Functionality (SOTIF) that aims to ensure that the ADS operate safely within their intended functionality. SOTIF focuses on preventing or mitigating potential hazards that may arise from the limitations or failures of the ADS, including hazards due to insufficiencies of specification, or performance insufficiencies, as well as foreseeable misuse of the intended functionality. However, the challenge lies in ensuring the safety of vehicles despite the limited availability of extensive and systematic literature on SOTIF. To address this challenge, a systematic literature review (SLR) on SOTIF for the ADS is performed following the preferred reporting items for systematic reviews and meta-analyses (PRISMA) guidelines. The objective is to methodically gather and analyze the existing literature on SOTIF. The major contributions of this paper are: (i) presenting a summary of the literature by synthesizing and organizing the collective findings, methodologies, and insights into distinct thematic groups, and (ii) summarizing and categorizing the acknowledged limitations based on data extracted from an SLR of 51 research papers published between 2018 and 2023. Furthermore, research gaps are determined, a comparative analysis of methods supporting SOTIF is provided, and supplementary insights from recent publications that address these gaps are presented. Based on the findings, future research directions are proposed.

## Keywords

Automated Driving System (ADS), Functional Safety (FuSa), Preferred Reporting Items for Systematic Reviews and Meta-Analyses (PRISMA), Safety of the Intended Functionality (SOTIF), Systematic Literature Review (SLR)

## 1. Introduction

Automated driving has emerged as a significant domain of research and development, with the potential to revolutionize the automobile industry. Automated driving involves an automated vehicle equipped with the ADS operating without human driver intervention. The automated vehicle performs driving tasks, including acceleration, braking, steering, and navigation, using perception sensors to build situational awareness and complex perception algorithms, with actuation performed by electrical and/or electronic (E/E) systems.

Safety is of paramount importance to prevent accidents that may occur during automated driving due to system failure or malfunction, fostered by the increasing complexity of the realization of the ADS. Given the complexity of an ADS and the associated deployment risks, understanding safety concerns necessitates a multidisciplinary approach that includes technical, ethical, and legal perspectives. Challenges arise during the design and implementation phases due to the lack of standardized conditions for specifying intended functionality. This necessitates ongoing adjustments to the specified functionality, based on an understanding of real-world system operations and variations in cultural usage [1].

The safety related to automated driving can be approached from two aspects: Functional Safety (FuSa) and Safety of the Intended Functionality (SOTIF). ISO 26262 [2] and ISO 21448 [3] are the international standards for FuSa and SOTIF, respectively. FuSa ensures that the system's inherent risk is reduced to an acceptable level, with respect to malfunctions of E/E components, while SOTIF ensures the system has an acceptable risk with respect to functional insufficiencies and performance limitations. In the following, the term system is used in place of ADS.

Machine learning (ML)-based systems are adopted in automated vehicles for tasks such as perception and planning. Despite their benefits, incorrect outputs from ML systems can compromise safety. ML-based systems process sensor data and make real-time decisions in unpredictable environments. They have inherent limitations, including biases in training data, incomplete datasets, and unexpected prediction errors. These issues can lead to performance insufficiencies and pose safety risks. For example, ML-based systems may react unpredictably to new events (i.e., lack of generalization) or behave inconsistently with similar inputs (i.e., lack of robustness). Optimal performance requires balancing task complexity, system capabilities, and data adequacy. Overly complex systems can be overfit to noise, while overly simple systems can exhibit bias by ignoring relevant relationships. Achieving this balance is crucial for reliable performance in automated vehicles [4].

ISO 21448 [3] provided guidelines to mitigate these risks through rigorous verification and validation, including extensive testing under varied conditions such as software-in-the-loop, hardware-in-the-loop, and vehicle-level testing. These guidelines help identify and mitigate risks associated with the intended functionality of the system, encompassing software and hardware failures, as well as challenging driving conditions caused by adverse weather or critical situations. The driving scenarios within the operational design domain (ODD) of the system are classified into four areas based on their potential to cause hazardous behavior.

- Known, not hazardous scenarios (Area 1)
- Known, hazardous scenarios (Area 2)
- Unknown, hazardous scenarios (Area 3)
- Unknown, not hazardous scenarios (Area 4)

The goal of SOTIF is to evaluate and reduce the risk of hazardous behavior in Area 2 and Area 3 by analyzing the intended functionality, modifying the system's functionality, and verifying and validating the system's performance [3].

SOTIF measures include modifying system functionality or sensor performance requirements, implementing redundancy and fail-safe mechanisms, and conducting thorough testing and validation. Ensuring SOTIF involves verifying the ADS's intended functionality is safe, considering system limitations and potential environmental and operational hazards. Successful SOTIF implementation means safety measures are effectively in place, working as intended to safeguard the ADS's functionality.

For instance, system modifications involve improving sensor performance or accuracy, while redundancy measures involve implementing additional sensors to ensure the intended functionality of the system in the event of a failure. Fail-safe mechanisms include emergency braking systems or other safety features that activate automatically in hazardous situations. These measures complement each other to ensure that the vehicle's intended functionality is maintained while minimizing the risk of hazardous behavior.

The significance of ensuring the safety of automated vehicles with regard to their intended functionality is highlighted in the study by Takacs et al. [5]. To ensure SOTIF, a holistic approach is required, involving the analysis of the system's intended functions, the identification of potential hazards and risks, and the development of appropriate mitigation strategies [6]. The study conducted by Zhu et al. [7] examined the technical challenges involved in ensuring SOTIF, encompassing the need for risk assessments, the importance of scenario-based testing, and the role of simulation and validation in verifying the safety and reliability of the system.

From 2020 to 2023, numerous publications have addressed SOTIF, including works by Birch et al. [6], Xu et al. [8], Hoss et al. [9], Zhao et al. [10], Wang et al. [4], and Saberi et al. [11], covering diverse topics from human–machine driving mode switch to formal methods beyond SOTIF. The above-mentioned publications were selected for their significant contributions within the specified timeframe, specifically addressing key aspects of human–machine interaction, formal methods, and scenario-based testing related to SOTIF.

Cao et al. [12] presented an analysis of automated vehicle localization in foggy conditions through SOTIF analysis and a three sigma-criterion-based adaptive extended Kalman filter (EKF). It proposes a functional modification strategy incorporating visibility recognition and adaptive filtering to mitigate SOTIF-related risks by improving system resilience against environmental uncertainties. The paper [13] introduces a method for creating ADS test scenarios using combinatorial testing and parameter sampling to balance exploration and scenario space utilization. It highlights challenges such as parameter discretion and expert knowledge dependence, proposing future improvements with real-world data and optimization.

Moreover, Zhang et al. [14] and Birkemeyer et al. [15] conducted SLR on finding critical scenarios and on scenario generation techniques for verifying and validating ADS in the context of the SOTIF but not directly on SOTIF. Consequently, this paper pioneers an SLR focused on SOTIF, adopting a structured PRISMA approach. Through this SLR, the factors associated with the successful implementation of SOTIF measures, the challenges that arise when ensuring SOTIF for an ADS, and research gaps from the existing literature on SOTIF have been determined. In addition to the SLR, this paper includes a comparative analysis of methods supporting SOTIF to evaluate their effectiveness, challenges, and practical applicability. Furthermore, supplementary insights from recent publications that directly address identified research gaps are presented, thereby extending the literature scope and providing the most up-to-date findings relevant to the topic.

An overview of the significance of ML-based systems for perception and situational awareness in automated vehicles is provided, but the primary focus is the SLR on SOTIF. Future research directions are proposed on addressing ML models for handling situational awareness scenarios, the implications of uncertainty in ML-based perception algorithms for SOTIF analysis, and developing frameworks for modeling SOTIF scenarios and generating test cases.

The limitations of this paper are twofold. First, the review focuses on literature published between 2020 and 2023. Although this timeframe is relatively recent, it is possible that relevant research published outside this period has not been included in the review. Second, the paper bases its conclusions on a limited number of research papers due to the exclusion criteria of the review methodology, which may restrict the scope of its findings.

### 1.1. Research Questions

Based on the objective to methodically gather and analyze the existing literature and to synthesize the current state of knowledge on SOTIF, the research questions (RQ) are formulated as follows:

- RQ1. What factors are associated with the successful implementation of SOTIF measures?
- RQ2. What safety challenges arise when ensuring SOTIF for ADS?
- RQ3. What are the research gaps determined from the SLR of existing literature on SOTIF?

### 1.2. Structure of the Paper

The subsequent sections of this paper are organized as follows: Section 2 describes the review methodology employed for the systematic review. Section 3, based on data extracted from 51 research papers, discusses the findings by presenting a literature summary, acknowledging limitations, and addressing RQ. Additionally, Section 3 includes a comparative analysis of methods supporting SOTIF in Subsection 3.4 and presents supplementary insights from recent publications that address identified research gaps in Subsection 3.5. Future research directions are proposed and briefly described in Section 4. Lastly, Section 5 concludes by summarizing the findings of this SLR.

## 2. Review Methodology

A systematic literature review (SLR) employs a structured approach to analyze and summarize the existing literature on a specific topic or research question. Its purpose is to identify research gaps, summarize current state of knowledge, and suggest future research directions [16]. The review methodology adheres to the Preferred Reporting Items for Systematic Reviews and Meta-Analyses (PRISMA) guidelines [17]. It includes four stages: (1) developing a search strategy, (2) defining exclusion criteria, (3) conducting the selection process (identification, screening, and quality assessment), and (4) data extraction. Figure 1 depicts the review methodology.

In the first stage of the review methodology, a search strategy is developed that details the databases considered in the review and the base string consisting of search terms used for inquiring the databases.

In the second stage of the review methodology, exclusion criteria are defined to exclude literature that does not meet specific requirements of the review.

The third stage of the review methodology encompassed three phases: identification, screening, and quality assessment (QA). In the identification phase, the search strategy is applied to the selected databases. Screening refines the eligible literature by reading the title and abstract of the publications, checking full-text accessibility, and rejecting those that meet the exclusion criteria. QA is then performed to evaluate methodological rigor and literature quality.

In the fourth stage of the review methodology, a data extraction is performed to collect pertinent information from each included publication, encompassing details on the research approach, results and findings, limitations

**FIGURE 1** Proposed review methodology. (Data taken from Refs. [18–21].)

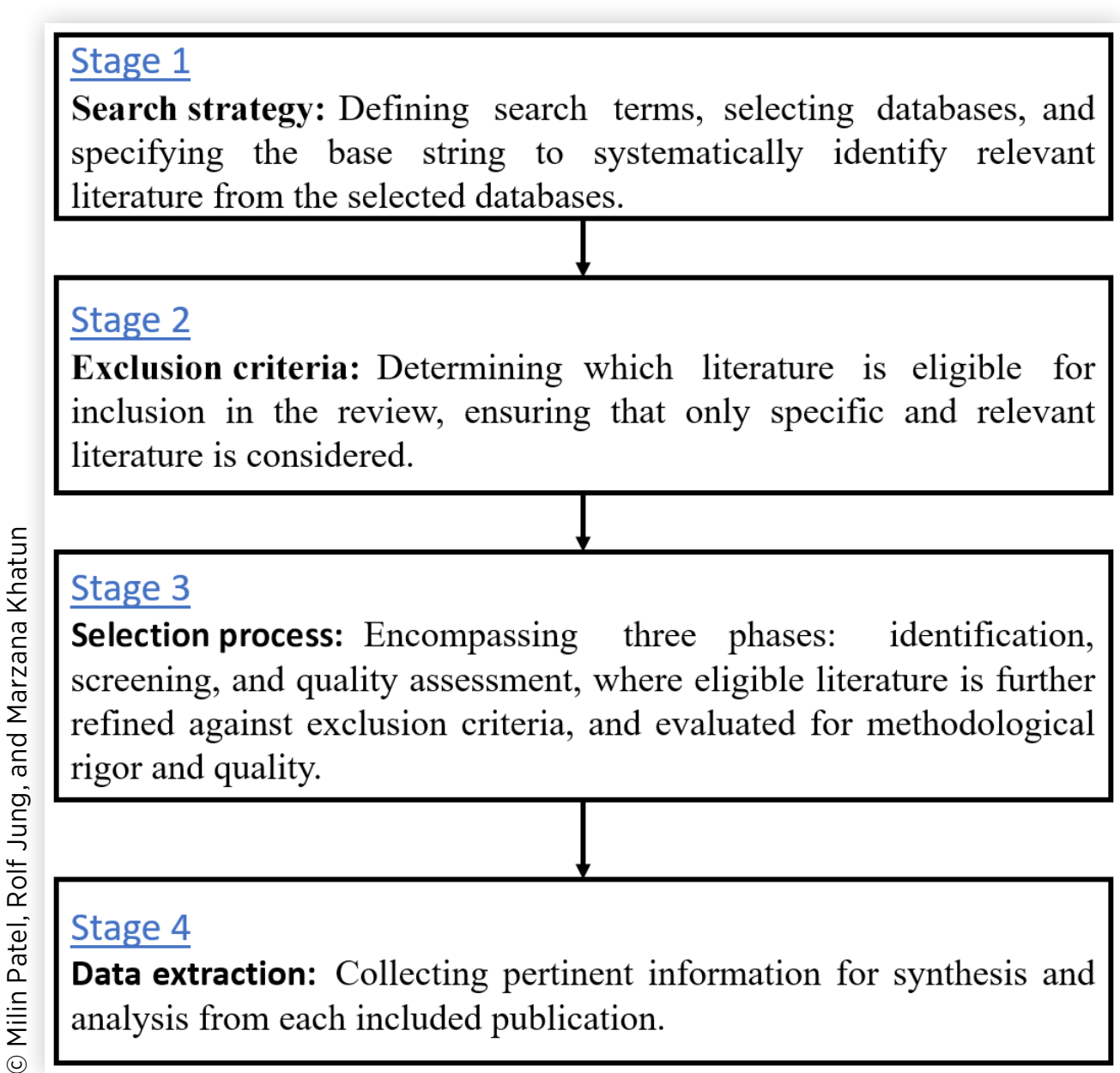

acknowledged by the authors, and conclusions and implications for synthesis and analysis.

Parsifal [22] is used as an online tool employed to assist in designing the review methodology, performing the selection process, and refining the selected literature against exclusion criteria including their QA.

## 2.1. Search Strategy

The search strategy entails the selection of databases and the formulation of a base string with relevant search terms. The defined search terms include *Safety of the Intended Functionality (SOTIF)*, *ISO 21448*, *automated driving*, and *Automated Driving System (ADS)*, which are combined using Boolean operators and truncation.

The base string formulated and utilized for the query is as follows:

*("Safety of the Intended Functionality (SOTIF)*" OR "ISO 21448") AND ("automated driving*" AND "Automated Driving System (ADS)*")*

The use of truncation in the base string by including a truncation symbol, typically an asterisk (*), enables different forms and grammatical variations of the search terms, thereby increasing the likelihood of retrieving relevant publications. In the provided base string, by including a truncation symbol (*) ensures the inclusion of the terms such as SOTIF-related, SOTIF-specific, automated driving, or any other forms and grammatical variations derived from *Safety of the Intended Functionality (SOTIF), automated driving*, and Automated Driving System (ADS)** in the search results.

Table 1 showcases the selected databases known for their established reputation as major repositories of academic publications and scientific literature in computer science, engineering, and transportation domains.

ACM Digital Library and IEEE Xplore provide access to academic publications including conference proceedings and journals, on automated vehicles and safety-critical systems. Dimensions covers multidisciplinary research, while arXiv specializes in pre-prints and publications related to artificial intelligence and ML. SAE Mobilus focuses on mobility engineering research [23]. Springer provides an extensive library of peer-reviewed journals and books, and external sources include patents, PhD theses, and other relevant publications, offering a diverse range of literature beyond the core databases.

**TABLE 1** Selected database.

| Database | URL |
|---|---|
| ACM Digital Library | https://dl.acm.org/ |
| IEEEXplore | https://ieeexplore.ieee.org |
| Dimensions | https://app.dimensions.ai/ |
| arXiv | https://arxiv.org/ |
| SAE Mobilus | https://saemobilus.sae.org |
| Springer | https://link.springer.com |
| External sources | N/A (Explanation provided below) |

Although external sources do not have specific URLs listed in the Table 1, they are valuable for accessing unpublished or specialized publications on SOTIF and the ADS. Despite the eventual exclusion of non-peer-reviewed preprints from arXiv in the QA phase, their initial consideration ensures no potentially impactful study is overlooked in the initial stages of the review.

Other well-known databases such as Web of Science, INSPEC, and Scopus were not included in the review for two main reasons. First, the selected databases (ACM Digital Library, IEEE Xplore, and SAE Mobilus) are more focused on research related to SOTIF and ADS, ensuring relevant content is retrieved. These repositories specialize in peer-reviewed studies on safety-critical systems and automated driving, making them suitable for the scope of this review. Second, limiting the number of databases helps to maintain a balance between relevance and search efficiency. While broader multidisciplinary databases such as Web of Science and Scopus provide valuable content, they often include research beyond the specific technological focus of this review. The exclusion of certain databases may omit insights from other fields; however, the inclusion of sources such as Dimensions and Springer compensates for this by covering a wide range of research relevant to SOTIF and ADS.

## 2.2. Exclusion Criteria

The exclusion criteria are defined to exclude the literature that does not meet the specific requirements of the review. Table 2 presents the exclusion criteria used to determine the eligible literature for the selection process.

E# column in Table 2 represents the distinct identifier for each exclusion criterion, labeled as E1 to E5. The statement column in Table 2 provides an explanation of each exclusion criterion along with references. The exclusion criteria E1 to E3 are applied during the identification phase to remove publications that do not meet specific requirements. Conversely, E4 and E5 are applied in the screening phase to exclude certain publications from further consideration. The goal is to ensure that only publications closely aligned with the research objectives are considered, while maintaining significance and quality in the review process.

## 2.3. Selection Process

The aim of the selection process is to gather relevant publications on SOTIF, assess their eligibility, and verify their quality and validity. The selection process is summarized in Figure 2. The PRISMA flow diagram is a graphical representation that outlines a number of records that are identified, screened, and included in this SLR review.

The selection process consists of three main phases: identification, screening, and QA.

**2.3.1. Identification** In the identification phase, the search strategy discussed in Subsection 2.1 is applied to gather publications from selected databases. The base string is adapted as necessary to retrieve relevant publications effectively.

**TABLE 2** Exclusion criteria for the selection process.

| E# | Criteria | Statement |
|---|---|---|
| E1 | Papers with titles that are not in English | The criteria are defined based on References [18–20]. |
| E2 | Papers that were published before 2018 | The standard for SOTIF (ISO 21448) was published in 2022. ISO/PAS 21448 was released in 2019. Therefore, by limiting the search to papers published from 2018 onward, the review can capture relevant and up-to-date literature on SOTIF. |
| E3 | Title of the papers that contain keywords that include proceedings, conference, symposium, workshop, or book | It aims to avoid indexing titles that represent events or collections rather than individual peer-reviewed research. This ensures a focus on substantive, peer-reviewed contributions directly relevant to SOTIF and ADS research. |
| E4 | Neither the title nor the abstract mentions at least one of the search terms | The search terms are: (i) Safety of the Intended Functionality (SOTIF), (ii) ISO 21448, (iii) automated driving, and (iv) automated driving system. The criteria are defined based on Reference [21]. |
| E5 | Full text of the papers is not accessible | The criteria are defined based on References [19, 20]. |

**FIGURE 2** PRISMA flow diagram for the selection process. (Data taken from Ref. [19].)

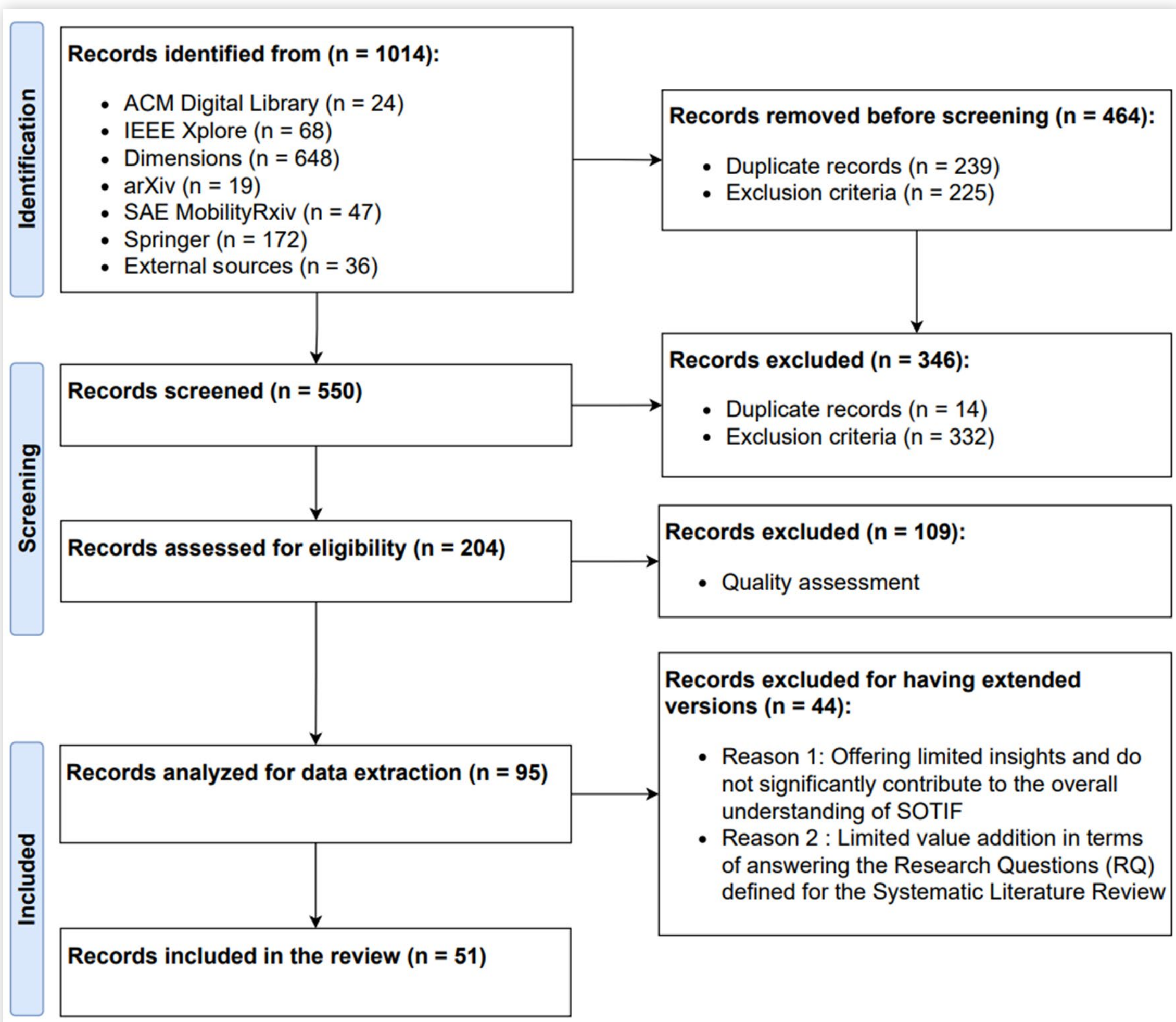

A total of 1014 search results were obtained. Duplicates arise when a publication is indexed in multiple databases or when multiple versions of a publication are available. A thorough removal of duplicates is performed to ensure each publication is counted only once in the SLR process.

A total of 239 duplicate records are excluded using the Parsifal tool [22]. The exclusion criteria, E1 to E3, detailed in Table 2, are then applied, leading to the exclusion of an additional 225 records.

**2.3.2. Screening** The screening phase review consists of reading the title and abstract of the publications and rejecting those that meet the exclusion criteria, E4 and E5, as outlined in Table 2. Initially rejected papers undergo another assessment to validate the exclusion.

During the screening phase, a total of 346 records were excluded from the initially identified 550 records. Duplicate records, identified by titles with invalid characters resulting from exporting search results from databases, led to the exclusion of 14 duplicate records.

**2.3.3. Quality Assessment** The included records from screening in this review follows the QA criteria presented in Table 3.

QA# column in Table 3 represents the distinct identifier for each exclusion criterion, labeled as QA1, QA2, and QA3. The statement column provides a detailed explanation of each exclusion criterion along with reasoning and references.

This review, conducted by three authors, employs a consensus-based approach, qualitatively assessing each criterion with a "Yes" or "No" response. Records receiving "Yes" responses for all QA criteria proceed to data extraction.

Figure 3 presents an overview of the QA responses for the 204 eligible records, categorizing responses into four groups: (i) all "Yes" responses, (ii) two "Yes" responses, (iii) one "Yes" response, and (iv) all "No" responses.

**TABLE 3** Quality assessment criteria in the form of questions.

| QA# | Criteria (questions) | Statement |
|---|---|---|
| QA1 | Is the paper published in a peer-reviewed journal or conference proceeding? | The paper published in a peer-reviewed journal or conference proceeding indicates that the paper has undergone rigorous evaluation and meets academic standards. |
| QA2 | Does the paper provide value for practice or research? | This criteria is defined based on reference [18].<br>By reading the abstract, introduction and/or conclusion, it can be determined whether the paper offers value for practice by providing practical insights, guidelines, or recommendations for industry application.<br>Similarly, a paper providing value for research contributes to the development of future research directions, methodologies, or frameworks. |
| QA3 | Does the paper provide adequate information regarding the context of SOTIF? | This criteria is defined based on reference [18].<br>This criterion evaluates if the paper covers SOTIF-related use cases, triggering conditions, foreseeable misuse, sensor performance limitations, machine learning challenges, and uncertainties, as outlined in ISO 21448. |

**FIGURE 3** Distribution of the quality evaluation scores obtained from assessing the eligible records considered in the review.

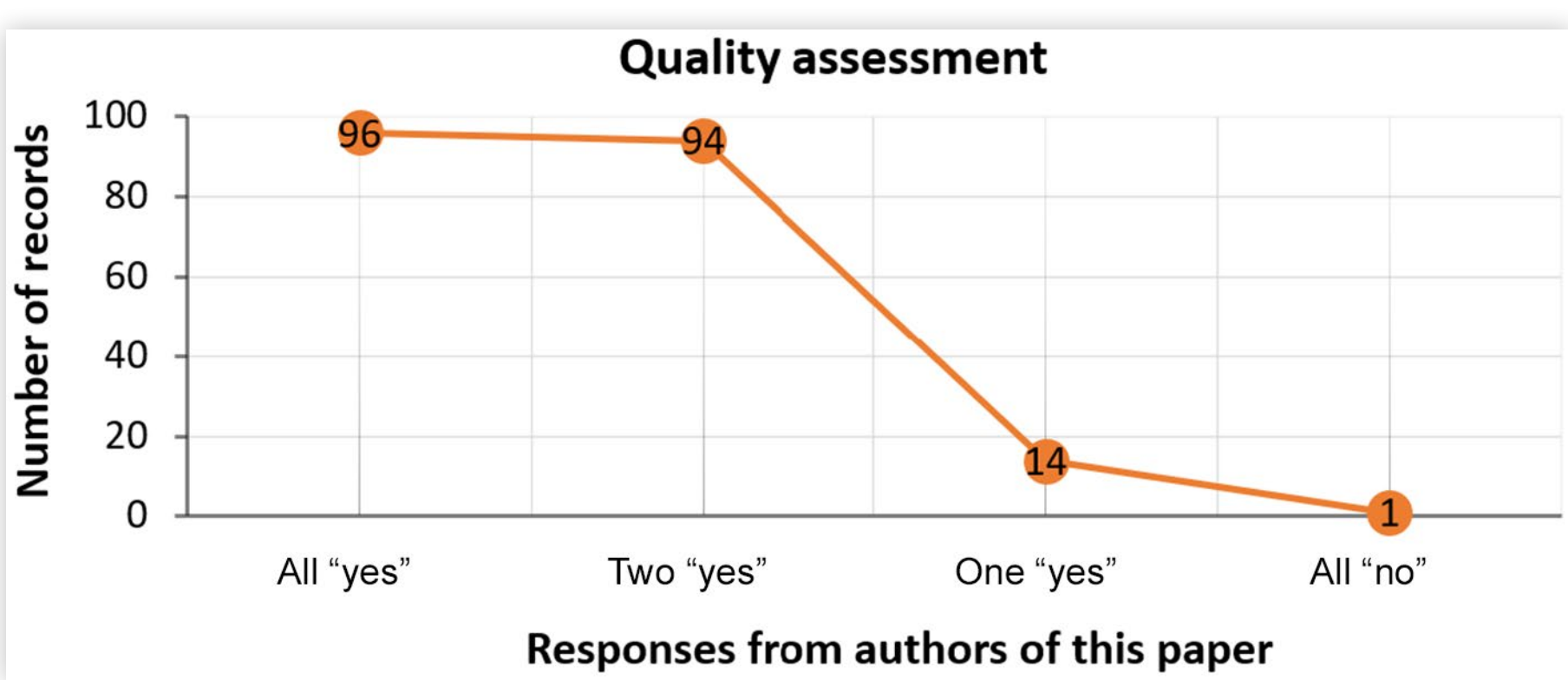

After applying the three QA criteria, 108 records are rejected. Consequently, the remaining 96 records are subjected to data extraction.

## 2.4. Data Extraction

The data extraction stage analyses records selected after the QA to extract pertinent information. This information is provided in Table 4 in the form of a list.

Although data extraction usually does not exclude records in a SLR, 44 out of the 96 analyzed records were identified as extended versions and excluded to maintain the singularity of the review. The exclusion was based on two objective factors: (1) limited insights or contribution to the overall understanding of SOTIF and (2) redundancy with other records that provided deeper insights into the RQ. By focusing on records that added distinct value, the review ensured a streamlined and focused synthesis. As a result, 51 records are included in this review.

**TABLE 4** Data extraction list.

| Data extraction list | Statement |
|---|---|
| Research approach | Describes the research approach, including overviews, methods, frameworks, architectures, methodologies, or concepts proposed or mentioned in the respective paper. |
| Results and findings | Summarizes the main findings and results related to SOTIF. |
| Limitations acknowledged by the author | Identifies any limitations or constraints acknowledged by the authors in their respective papers, which may affect the validity or generalization of the findings. |
| Conclusions and implications | Key conclusions and implications of the paper for the field of SOTIF and the ADS. |

Furthermore, the summary of records excluded from the review is depicted in Figure 4, which illustrates the number of records that were excluded for each specific criterion, as outlined in Table 2.

# 3. Discussion

This section is divided into five subsections: literature summary (3.1), limitations acknowledged (3.2), addressing RQ (3.3), comparative analysis of methods supporting SOTIF (3.4), and supplementary insights (3.5). Each subsection provides a detailed discussion based on the data extracted from 51 included research papers, except for the supplementary insights, which include recent publications beyond the initial review.

## 3.1. Literature Summary

This subsection synthesizes and organizes the research approaches, findings, and conclusions from the included studies. The literature is summarized into distinct thematic groups encompassing holistic safety approaches, addressing uncertainties, hazard identification and risk assessment, safety requirements for ML-based perception, and scenario-based SOTIF analysis.

**3.1.1. Holistic Approach to Safety in Automated Driving** The research conducted by Kirovskii and Gorelov [24], Collin et al. [25], and Kinalzyk [26] collectively recommended a holistic approach that combines both functional safety (FuSa) and SOTIF.

Kirovskii and Gorelov [24] emphasized the importance of addressing hazards related to SOTIF in the ADS and argued that FuSa does not address the risks associated with non-deterministic parts and algorithms, including those used in ML. Collin et al. [25] acknowledged that SOTIF is not a complete approach for the safety and validation of the ADS and should be used in conjunction with FuSa. Collin et al. [25] proposed the rulebooks framework as a practical and flexible approach to specifying driving behavior that can help identify specific hazardous scenarios and guide behavior testing toward critical scenarios for validation.

**FIGURE 4** Distribution of excluded records by criterion.

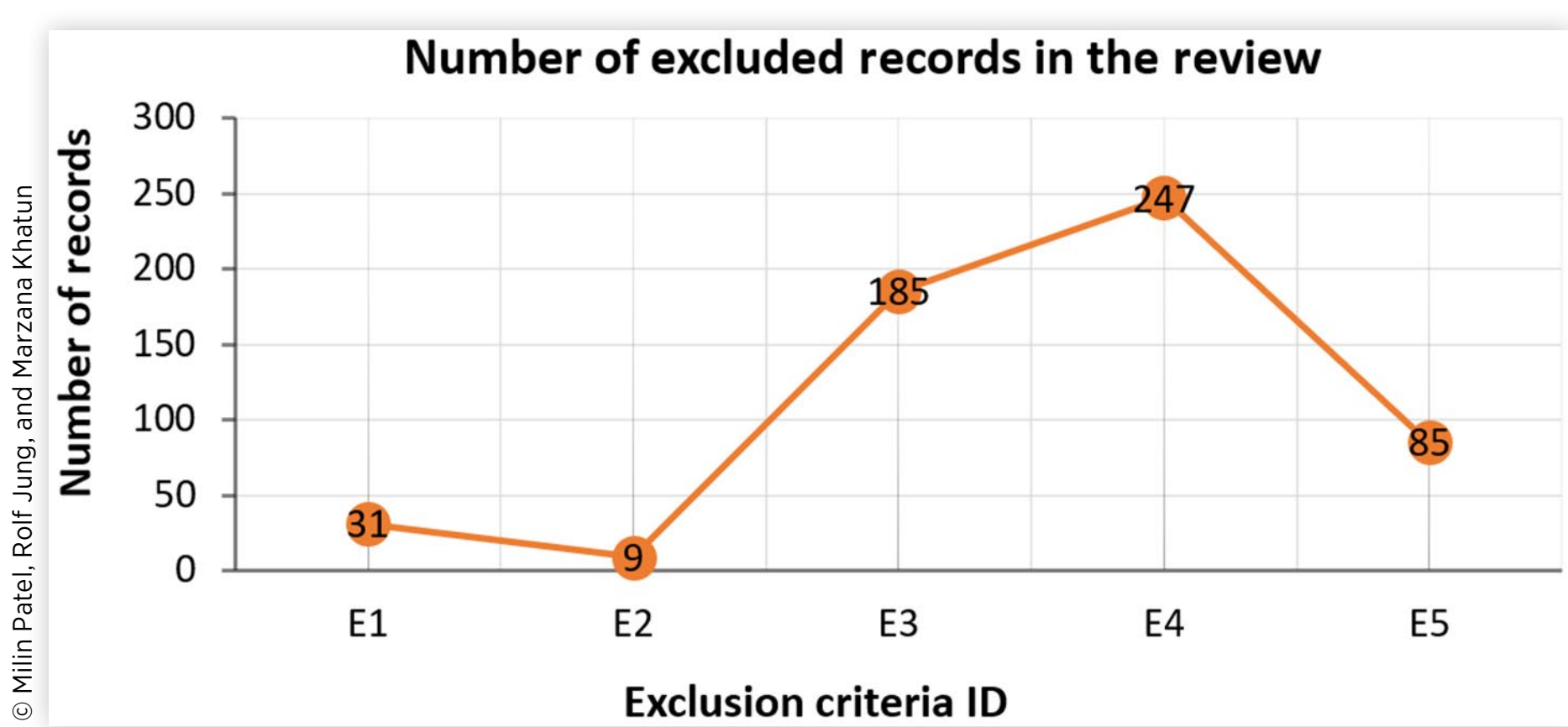

Kinalzyk [26] introduced a data-driven test management and optimization loop that integrates FuSa and SOTIF to address residual risks in highly automated driving. This approach focused on continuous testing and improvement to ensure safety across diverse system conditions. Building on this, [27] highlighted that FuSa and SOTIF cover only a subset of the potential risks in complex systems, particularly those involving sensor limitations and ML uncertainties. To mitigate these risks, the system view–based approach to validation (sys²val) framework [27] is proposed, aligning with the combined FuSa–SOTIF strategy proposed by Kirovskii and Gorelov [24].

Furthermore, [28] discussed a modular safety system design for intelligent autonomous vehicles, emphasizing that existing standards do not fully address all safety risks. Feth et al. [29] emphasized the need for a multi-aspect safety engineering approach for highly automated driving, integrating FuSa, functional insufficiencies addressed by SOTIF, and a novel focus on safe nominal behavior specification.

#### 3.1.2. Addressing Uncertainties in SOTIF

ISO 21448 [3] provided guidance on identifying and managing uncertainties, including those related to the external and internal vehicle environment, and the lack of robustness of the function, system, or algorithm with respect to sensor input variations, heuristics used for fusion, or diverse environmental conditions. It involves handling unforeseen challenges and risks arising from factors such as unpredictable driver behavior, environmental conditions, and operational complexities of the ADS.

There is a need to address the challenges associated with unpredictable driver behavior and accidents that are not caused by component failure but by situations that were not planned for during development [30]. The uncertainties referred to in paper [30] relate to the velocities of traffic participants and are influenced by various factors, including weather conditions, visibility, and traffic volume. Lotto et al. [30] discussed the challenges associated with uncertainties and propose a copula-based approach for modeling stochastic data, which can be used to extract influential stochastic parameters from real-world measurements and real-time data.

The uncertainties addressed in paper [31] are related to the operational environment, sensing, understanding the environment, and the complexity of the system. Adee et al. [31] proposed a novel methodology using extended evidential networks (EEN) that quantifies uncertainties arising from randomness, lack of knowledge, and a state of complete ignorance. Uncertainty, mentioned in paper [32], refers to the lack of confidence or reliability in ML-based object detection algorithms for automated vehicles when operating in scenarios affected by environmental conditions, including but not limited to extreme weather and adverse lighting conditions.

The detection of out-of-distribution (OoD) scenarios [33] to manage uncertainties arising from ML failures supports the argument that uncertainty in object detection algorithms can be mitigated by integrating ML-based methods with a robust safety lifecycle. This extends the discussion on how uncertainties related to sensing and understanding the environment, as in Lotto et al. [30], can be better addressed using novel ML techniques.

Another source of uncertainty lies in the data used for critical analyses. The reliance on the GIDAS database, which is geographically limited to Germany, raises concerns about the representativeness of the data when applied to broader contexts [34]. Finally, [35] highlighted the importance of identifying and quantifying hazardous scenarios that might not be captured in existing databases, emphasizing the need for systematic methodologies to address rare but high-consequence events during the validation of ADS.

#### 3.1.3. Hazard Identification and Risk Assessment for SOTIF

Identifying potential hazards associated with the intended functionality of the automated vehicle involves assessing the likelihood and severity of associated risks and defining acceptance criteria to ensure that the risks are reduced to a level that is considered acceptable.

Numerous methodologies for hazard identification and analysis, including system-theoretic process analysis (STPA)–based approaches [36] and Bayesian network analysis [37], are investigated.

Zhang et al. [36] and Delaus et al. [37] developed a methodology that extends the STPA framework by incorporating a causal scenario classification system and a complex network-based evaluation for analyzing SOTIF-related hazardous factors. This methodology evaluates the interaction complexities and performance limitations of Intelligent Railway Driving Assistance Systems (IRDAS). Delaus et al. [37] introduced Bayesian networks to model failure-event relationships quantitatively, providing a systematic and complementary approach to SOTIF analysis. Furthermore, Qin et al. [38] provided technical guidance and recommendations for analyzing the safety of autopilot functions in automated vehicles, which contributes to the development of methodologies for evaluating potential hazardous behaviors in known and unknown hazardous scenarios.

Diverse strategies for SOTIF risk mitigation are proposed, including robust non-fragile fault-tolerant control (RNFTC) strategy [39] and a framework that uses ML and multiple sensors [40]. The proposed RNFTC strategy takes into account various forms of uncertainties, including system uncertainty, perception and actuation performance limitations, and multi-source disturbances, as well as controller perturbations [39]. The proposed framework monitors and mitigates safety risks caused by vehicles exceeding ODD constraints in complex traffic scenarios by monitoring weather conditions, vehicle behavior, and road conditions [40].

Kramer et al. [41] proposed an integrated method for hazard identification and risk assessment that combines established approaches, which are fault tree analysis (FTA) and event tree analysis (ETA), and extends them to enable their applicability to the SOTIF for the ADS. The quantitative risk norm (QRN) approach addresses gaps in existing risk assessment methodologies by providing a quantitative method for hazard analysis in ADS [42].

A research on SOTIF for LKA [43] highlighted the use of STPA to systematically identify hazardous control actions and evaluate functional insufficiencies under diverse triggering conditions. The study emphasized how data-driven methodologies can extend the utility of the SOTIF standard for the LKA function, accounting for unexpected behaviors and environmental factors.

**3.1.4. Safety Requirements for ML-Based Perception** Safety requirements refer to specifying the relevant functionalities and use cases, identifying and mitigating the limitations of ML algorithms, evaluating the safety of ML-based perception systems through appropriate testing, and using appropriate data collection processes to minimize biases and other limitations.

Celik et al. [44] presented an STPA-based approach for eliciting safety requirements for ML-based perception components. It enables the linkage between the derived SOTIF requirements to the properties of ML components like performance and provides a set of safety requirements that address SOTIF-related triggering conditions associated with performance limitations of ML-based algorithms. Borg et al. [45] suggested that the assurance of machine learning for use in autonomous systems (AMLAS) and SOTIF frameworks can support the safety assurance of ML-based systems in automated driving and provide practical insights, guidelines, and recommendations that can be directly applied in the industry.

Kaneko et al. [46] suggested a safety analysis framework for deep neural network (DNN) systems in automated driving that models the entire system, using system-theoretic accident model and processes (STAMP) and analyses interactions with other system elements using STPA and causal analysis based on system theory (CAST) analysis methods. Hacker and Seewig [47] proposed an insufficiency-driven DNN error detection approach on traffic sign recognition use case by addressing functional insufficiencies and triggering conditions.

The limitations of self-aware trajectory prediction, particularly the assumption of prediction module errors, provide insights into the safety requirements for ML-based systems, especially object detection algorithms [48]. This reinforces the need for continuous refinement in ML-based perception systems, as mentioned by Celik et al. [44]. Furthermore, ensuring hardware compliance with FuSa and SOTIF for autonomous driving systems contributes to the safety analysis of DNNs, as discussed by Kaneko et al. [48], and highlights the need for robust hardware-software integration within the ML development process [49].

**3.1.5. Scenario-Based SOTIF Analysis** Scenario-based analysis refers to the process of identifying and analyzing potential hazardous scenarios that may arise when the driving conditions exceed the performance or functional limitations of one or more system components or from human factor considerations. The scenarios are derived through a structured framework that combines variables derived from different sources, including use cases, triggering conditions, foreseeable misuse, and limitations of the sensors.

The significance of a SOTIF-related scenario dataset that covers different weather, seasons, and times of the day, and includes trigger conditions that can significantly degrade the perception ability and uncertainty estimation methods for perception-oriented detection algorithms, is highlighted by [50]. Peng et al. [32] used a SOTIF-related scenario dataset to evaluate ML-based object detection algorithms, highlighting the importance of considering scenarios beyond the intended use of the system.

Maier et al. [51] proposed a novel approach to ensure the safety of automated vehicles using causal models and causal metrics. This approach can identify critical parameter configurations and help address the difficulty in creating a sufficient set of scenarios for SOTIF. Moreover, Meyer et al. [52] introduced a systems modeling language (SysML)-based approach, designed to ensure SOTIF compliance, by integrating scenario and use case analysis through structured identification and mitigation of risks. Validated by an autonomous parking case study, this approach bridges system requirements with safety analysis, demonstrating practical applicability.

The optimization and validation method for scenario-based safety analysis focuses on reducing the number of scenarios while maintaining safety standards [53]. This aligns with the SOTIF analysis framework by Meyer et al. [42], emphasizing the need for structured validation and prioritization. Additionally, evaluating scenario distribution for verifying and validating SOTIF systems is essential for maintaining system reliability, as discussed by Peng et al. [54], contributing to the overall process of safety validation [55]. Finally, [54] emphasized the importance of simulation-based testing and evidence-driven development feedback loops to ensure ADS safety and reliability.

## 3.2. Limitations Acknowledged

This subsection presents the limitations acknowledged by the authors in their respective research papers, summarizing them into categories such as limitations in validation frameworks, scope and applicability of proposed approaches, data quality and simulation challenges, and human factors and ethical considerations.

**3.2.1. Limitations in Validation Frameworks** The proposed unified evaluation framework for automated vehicles by Roshdi et al. [56] is considered as proof of concept and necessitates real-world validation. Additionally, the rulebooks framework proposed by Collin et al. [25] for ensuring the safety of intended driving behavior is not a generic solution and should be used alongside other validation methods. The methodology proposed by Kaiser et al. [57] for integrating safety analyses and simulation in autonomous emergency braking system (AEBS) use cases necessitates validation in real-world scenarios to ensure its effectiveness and practicality. This is because the proposed methodology is based on a single use case and may not be directly applicable to other systems, and the challenges associated with implementing the methodology in practice need to be addressed.

Moreover, an approach for SOTIF risk mitigation based on unified ODD monitoring proposed by Yu et al. [40] has some limitations. First, the proposed framework has not been tested in real-world scenarios, and further testing is required to validate its effectiveness in real-world environments. Second, the proposed framework relies on accurate sensor data, and any errors or malfunctions in the sensors could affect the performance of the framework [40].

The above-mentioned limitations emphasize the need for refinement of validation frameworks to ensure their applicability and effectiveness.

#### 3.2.2. Scope and Applicability Limitations of Proposed Approaches

Putze et al. [58] and Waschle et al. [59] acknowledged limitations in the scope and applicability of their proposed approaches. The approach for quantification of SOTIF validation of the ADS is based on a simplified model of the lane-keeping system, potentially limiting its direct applicability to other ADS [58]. Similarly, a study on artificial intelligence (AI) safety in highly automated driving acknowledges its focus only on highly automated driving and highlights potential limitations in generalizability to other domains [59].

Adee et al. [60] proposed a methodology for utilizing Bayesian networks to model triggering conditions and performance limitations in a scene to assess the SOTIF. However, the scope of the methodology is limited to the scene model and does not consider the dynamic nature of the environment, and relies on expert knowledge and real-world data, which may not be available or applicable in all driving scenarios.

The influence of fog weather on automotive vision target detection is studied for only fog weather conditions, but not extended to other weather conditions [61]. Feth et al. [29] acknowledged that their proposed safety engineering methodology for highly automated driving is in the early stages of development and that further research is needed to evaluate its effectiveness and applicability to different use cases.

The scope and applicability limitations underscore the necessity of developing adaptable methodologies that can address diverse situations, environments, or conditions in which it may be applied. The situations may differ in terms of specific requirements, constraints, objectives, and may require adaptations in proposed approaches.

A review of risk assessment methodologies highlights their limited applicability across different contexts, further supporting the need for adaptable approaches [62]. The controlled testing environment of the Quadsight® vision system similarly restricts its generalizability to real-world scenarios, emphasizing the necessity of broader testing conditions [63]. The review of FuSa design methodologies lacks empirical evidence and case studies, which supports the need for further validation and testing to ensure applicability across different autonomous vehicle systems [64]. Additionally, the limitations in autopilot functionality and evaluation standards underscore the need for detailed scenario collections for reliable human–machine interaction testing [8, 59]. Reliance on expert knowledge introduces variability, limiting the applicability of methods across different scenarios, as highlighted in approaches like those discussed by Feth et al. [36], where effectiveness varies depending on the context [65].

#### 3.2.3. Data Quality and Simulation Challenges

Peng et al. [32] and Borg et al. [45] recognized challenges related to data quality and simulation. Data quality refers to the accuracy, completeness, and reliability of the dataset used for analysis. The research paper by Peng et al. [32] on uncertainty evaluation of object detection algorithms acknowledged limitations in the size and diversity of the used dataset, as well as complexities in uncertainty analysis. The limitations of using simulation for empirical evaluations are highlighted, with emphasis on potential disparities between simulation and real-world scenarios [45].

Kirovskii and Gorelov [24] proposed a safety case model for ML that aims to prove the residual risk associated with functional insufficiencies in the object detection and classification function is acceptable. However, the proposed safety case model relies on data, including driving and accident statistics, as well as relevant triggering conditions. It is assumed that, if this data is not available or of poor quality, the safety case may not fully capture all possible driving scenarios and risks associated with the system.

Peng et al. [50] acknowledged that the proposed dataset may not cover all possible long-tail traffic scenarios, and the evaluation protocol may not be sufficient for safety-critical applications. Furthermore, the German In-Depth Accident Study database is used for criticality analysis of the ADS, but its limitations to Germany mean that the findings may not be representative for other regions or countries [66]. Challenges in validating ADS functions, particularly due to the complexity and variability of scenarios, highlight the need for robust simulation frameworks and realistic sensor models [67].

#### 3.2.4. Human Factors and Ethical Considerations

Human factors' considerations include designing the driver–vehicle interfaces to be intuitive and easy to use, minimizing driver distraction, and ensuring that the automated vehicle is ergonomically designed. Ethical considerations refer to the moral principles that guide decision-making and behavior of the ADS. Ethical challenges arise when the ADS are required to make decisions in complex and potentially hazardous situations. Resolving the ethical dilemmas and ensuring that the system's decision-making aligns with societal norms and values is a complex task.

Ethical considerations related to the allocation of control authority between the driver and the ADS can be complex and require careful consideration [68]. Yan et al. [68] mentioned that the uncertainty and variability of driver behavior make it challenging to design effective safety evaluation strategies and dynamic evaluation models of driver error.

The unified evaluation framework proposed by Roshdi et al. [56] for automated vehicles overlooked the impact of human factors on the ADS safety. Moreover, a research paper on rethinking certification for higher trust and ethical safeguarding of the ADS by Kusnirakova et al. [69] acknowledged the absence of comprehensive requirements for ethical considerations.

## 3.3. Addressing RQ

Based on the insights from Subsections 3.1 and 3.2, this subsection addresses the RQ by identifying factors associated with the successful implementation of SOTIF measures, safety challenges that arise when ensuring SOTIF for ADS, and research gaps determined from the SLR.

### 3.3.1. RQ1: What Factors Are Associated with the Successful Implementation of SOTIF Measures?

- **Risk Assessment and Risk Mitigation**

  The process of identifying, analyzing, and evaluating potential risks associated with the functionality of an ADS and implementing measures to reduce or mitigate these risks to ensure safety. The effectiveness of these measures is evaluated during the verification and validation phases of the SOTIF process to ensure that they are effective in reducing the identified risks [29, 36, 37, 39, 40, 42].
- **System Design**

  The systematic process of defining the architecture, components, interfaces, and data for an ADS to meet specified requirements, including the allocation of functionality to ML-based algorithms while considering safety aspects [24–26].

  System design includes ensuring that the system is designed to be safe and reliable, and that it can detect and respond to potential hazards. The system should be designed to be transparent, so that users can understand how it works and what its limitations are. It means that there must be clear and concise information about the system's capabilities and limitations, as well as any potential hazards or risks associated with its use.
- **Testing and Validation**

  The activities performed to verify and validate the performance and safety of ML-based components in an ADS, including testing under various conditions to ensure that the system functions as intended and meets safety requirements [32, 51, 67].

### 3.3.2. RQ2: What Safety Challenges Arise When Ensuring SOTIF for the ADS?

- **Unknown Scenarios**

  Situations or conditions that have not been encountered or anticipated during the development and testing of an ADS, posing challenges in ensuring the system's safe operation in novel or unexpected circumstances [30, 70].
- **Human Factors**

  Factors related to human interaction with the ADS, including user interfaces, communication protocols, and decision-making processes, which impact the safety and effectiveness of the system in real-world driving scenarios [68].
- **Sensor Limitations**

  Constraints or insufficiencies in the performance of sensors used in the ADS, such as inaccuracies, limited range, or susceptibility to environmental conditions, which can affect the system's ability to perceive and react to its surroundings accurately [32, 45, 71].
- **System Complexity**

  The degree of intricacy and interdependence of components, algorithms, and functionalities within an ADS, which can introduce safety challenges related to system integration, performance optimization, and fault tolerance in complex driving scenarios [25].

### 3.3.3. RQ3: What Are the Research Gaps Determined from the SLR of Existing Literature on SOTIF?

- **Limited Focus on Human Factors**

  Human driver behavior and interaction with the ADS can have a significant impact on the safety of the system, but there is a lack of understanding of how to effectively integrate human factor considerations into SOTIF methodologies.
- **No unified evaluation criteria for SOTIF measures**

  Unified evaluation criteria are necessary because they provide a consistent and objective way to evaluate the effectiveness of SOTIF measures. Without unified criteria, different stakeholders may use different evaluation methods, leading to inconsistent and potentially unreliable results. This can result in increased safety risks and potential harm to road users [35].

## 3.4. Comparative Analysis of Methods Supporting SOTIF

This subsection presents a comparative analysis of the methods and frameworks identified in the SLR review that support the SOTIF in ADS. The objective is to evaluate their effectiveness, challenges, and practical applicability. Table 5 summarizes these approaches, highlighting their individual contributions and interrelationships. By synthesizing these insights, practical implications are identified, findings are compared with existing literature, and recommendations are provided for their application.

The comparative analysis underscores the necessity of a multifaceted approach to address the safety challenges associated with SOTIF in ADS. The methodologies identified in the literature address different aspects of these challenges, including risk mitigation, uncertainty management, scenario-based validation, ML safety assurance, and real-time control and monitoring.

**TABLE 5** Comparative analysis of approaches and methodologies supporting SOTIF.

| Methodology/ approach | Description and insights | Considerations/challenges | References |
|---|---|---|---|
| Holistic approach combining FuSa and SOTIF | Integrates FuSa and SOTIF standards to address systematic failures and functional insufficiencies in ADS. Implements SOTIF requirements within the FuSa lifecycle, ensuring safety analysis throughout development and operation.<br>Highlights the necessity of addressing both malfunctioning behavior and performance limitations, essential in systems where AI significantly impacts safety. | Managing dual safety methodologies increases complexity.<br>Requires extensive documentation and rigorous lifecycle management to balance FuSa and SOTIF processes. | [24, 26] |
| Rulebooks framework | Specifies AV driving behavior through formalized, prioritized rules, aligning with SOTIF requirements. Scenario-agnostic and focuses on behavior, providing flexibility across various vehicle platforms and algorithms. | Defining a coherent rule set and priority structure is complex.<br>Requires resource-intensive testing and validation in real-world and simulated environments. | [25] |
| Integrated modular safety system (IMSS) and sys²val framework | Integrates hazard and risk analysis across hardware failures, environmental perception, human interaction, and AI functionality.<br>Emphasizes continuous validation and integration of FuSa and SOTIF in autonomous vehicles. | Increases system complexity due to aligning multiple safety requirements.<br>Requires substantial resources for ongoing validation in unpredictable environments. | [27, 28] |
| STPA-based SOTIF analysis | Applies STPA to identify unsafe control actions and causal factors, integrating them into the safety lifecycle to address functional insufficiencies.<br>Supports deriving safety requirements linked to hazards from ML limitations, component interactions, and system behaviors. | Requires expertise in modeling and analyzing control structures. Complexity in aligning with ISO standards.<br>Depends on detailed data for functional insufficiencies and operational scenarios. | [36, 43, 44] |
| Bayesian network analysis | Uses Bayesian networks to model and quantify uncertainties in system performance, supporting systematic safety evaluation.<br>Identifies weaknesses and calculates hazard rates across various configurations. | Complex modeling and interpretation of extensive data required for network construction and calibration.<br>Challenges in managing uncertainties when data is incomplete or estimated. | [37, 60] |
| Quantitative risk norm (QRN) | Tailors Hazard Analysis and Risk Assessment (HARA) for ADS by establishing quantitative risk norms.<br>Defines acceptable incident frequencies for different severity classes, deriving safety goals based on incident classification rather than listing all operational scenarios. | Ethical considerations in allocating incident frequencies.<br>Implementation complexities.<br>Significant effort required to quantify and validate incident frequencies, especially for complex ADS designs. | [42] |
| Probabilistic modeling and copula-based modeling | Handles uncertainties using probabilistic models. EEN and Bayesian networks model various uncertainties.<br>Copula-based models capture dependencies between driving parameters using real-world data, improving simulation reliability. | Require expertise in probabilistic reasoning.<br>Depend on data availability and quality.<br>Complex implementation and interpretation. | [30, 31, 60] |
| Out-of-distribution detection | Distinguishes between in-distribution and out-of-distribution data to identify gaps in training datasets and detect uncertain conditions during operation.<br>Supports the safety lifecycle of ML-based ADS by highlighting underrepresented areas and triggering fallback mechanisms. | Requires precise tuning of detection thresholds.<br>Relies on training data quality.<br>Integration with runtime safety mechanisms can be resource intensive. | [33] |
| Causal model-based engineering and scenario-based systems engineering | Utilizes structural causal models to represent causal relationships in automotive safety.<br>Facilitates scenario-based testing by identifying critical scenarios through probabilistic reasoning and counterfactual analysis.<br>Integrates scenarios into system engineering, enabling traceability and supporting SOTIF analysis. | Requires significant domain expertise.<br>Challenges in managing complexity, ensuring model consistency, and validation.<br>Iterative development and integration of diverse knowledge sources needed. | [14, 51, 52] |

(*Continued*)

**TABLE 5 (Continued)** Comparative analysis of approaches and methodologies supporting SOTIF.

| Methodology/ approach | Description and insights | Considerations/challenges | References |
|---|---|---|---|
| Robust non-fragile fault-tolerant control (RNFTC) and unified ODD monitoring framework | Develops control strategies and monitoring frameworks to ensure SOTIF compliance.<br>RNFTC mitigates risks from performance limitations, environmental disturbances, and system uncertainties.<br>Unified ODD monitoring framework monitors weather, vehicle behavior, and road conditions. | Requires complex control designs and advanced estimation techniques.<br>Challenges in integrating heterogeneous sensor inputs and adapting to dynamic scenarios. | [39, 40] |
| Assurance of machine learning components (AMLAS) | Provides a framework for developing safety cases for ML-based systems. Integrates safety mechanisms like out-of-distribution input rejection.<br>Relies on systematic testing and validation within a minimal ODD. | Challenges in ensuring sufficient safety evidence for ML-based systems.<br>Managing limitations in training data, especially from simulated environments. | [45] |
| Perception SOTIF evaluation frameworks | Focuses on evaluating and addressing perception-related SOTIF challenges using specialized datasets emphasizing long-tail traffic scenarios.<br>Utilizes methods estimating detection uncertainty to identify and mitigate perception insufficiencies. | Requires continuous dataset updates for broader coverage.<br>High computational resources needed.<br>Does not fully mitigate all uncertainties in perception algorithms or datasets. | [32, 50] |
| Self-surveillance and self-adaptation system | Provides a framework for real-time monitoring and mitigation of SOTIF risks.<br>Combines inherent risks from algorithm performance limitations and external collision risks.<br>Uses uncertainty estimation methods like Deep Ensembles and entropy quantification. | High computational demands for real-time risk quantification.<br>Requires accurate modeling of uncertainties.<br>Increases system complexity when integrating multiple AI modules. | [72] |
| SOTIF scenario hierarchy and metrics-driven scenario generation | Generates diverse test scenarios based on risk assessment and metrics like Scenario Potential Risk.<br>Uses genetic algorithms for scenario generation, enabling identification of critical cases. | Requires accurate risk quantification through probabilistic methods.<br>Managing computational complexity in scenario generation.<br>Balancing diversity and criticality in test cases is intricate. | [73–75] |

Holistic frameworks integrating FuSa and SOTIF provide safety coverage but introduce lifecycle complexity that requires efficient management strategies. Uncertainty management techniques, exemplified by probabilistic modeling and Bayesian networks, are crucial for handling data variability and sensor limitations but depend heavily on detailed and high-quality data. Scenario-based validation frameworks are essential for testing under diverse and critical conditions, capturing edge cases, but are computationally intensive and require iterative refinement.

ML safety assurance frameworks focus on ensuring that ML components meet safety standards, which is particularly important for ADS systems relying heavily on ML. However, limitations in training data and evidence generation present significant challenges. Real-time control and monitoring approaches address performance limitations and external disturbances in dynamic environments but may face scalability issues due to complexity and computational demands.

Comparison with existing literature shows a consensus on the need for integrated methodologies that address both functional insufficiencies and performance limitations inherent in ADS. Nevertheless, gaps exist in the practical applicability and scalability of these methods, particularly regarding data availability and computational resources. Existing studies often emphasize the effectiveness of individual approaches without providing solutions that are readily implementable in real-world systems.

Future research should focus on developing adaptable and scalable methodologies that can function effectively with limited data and computational resources. Addressing human factors and real-world applicability is crucial to bridge current gaps. Establishing unified evaluation criteria for SOTIF measures would improve consistency and comparability across different studies and applications.

Selecting appropriate methodologies should align with the specific needs and constraints of the ADS under development. Systems operating in highly uncertain environments benefit from uncertainty management techniques, while those heavily reliant on ML components should prioritize safety assurance frameworks. Often, a combination of methods is necessary to address SOTIF challenges. Implementing SOTIF measures effectively in ADS requires an approach that integrates various methodologies tailored to specific challenges. Careful selection and combination of appropriate methods can improve the safety and reliability of ADS, facilitating their wider acceptance and deployment. This comparative analysis provides a foundation for informed decision-making regarding the most suitable methodologies for particular aspects of SOTIF, aiding researchers and practitioners in advancing the field.

## 3.5. Supplementary Insights: Extending the Literature Scope

Recent publications from 2024 directly address specific research gaps identified in the SLR. These studies were not subjected to the PRISMA-guided systematic review but are included due to their direct relevance to the identified gaps. Although not evaluated under the same criteria, they provide valuable findings in hazard identification, scenario-based testing, handling of functional insufficiencies, integration of human factors, and validation methodologies related to SOTIF. Therefore, they are presented here as supplementary insights to complement the findings of the review.

A methodology combining hierarchical Bayesian networks with noisy gates to assess both known and unknown hazards in ADS has been proposed [76]. By integrating real-time data from the Internet of Vehicles, this approach dynamically updates hazard probabilities, enhancing adaptability to unforeseen conditions. This method aligns with ISO 21448's focus on unknown scenarios and performance insufficiencies.

An adaptive mining framework using a multi-population genetic algorithm to identify critical failure scenarios in ADS has been introduced [77]. By modeling scenarios using genetic algorithms and applying multi-dimensional fitness functions, the framework efficiently uncovers rare but high-risk events. This approach supports systematic hazard identification and risk mitigation, essential for SOTIF compliance. Challenges include computational complexity and the need for real-world validation to ensure applicability beyond simulated environments.

A scenario generation technique using Digital Twin technology to create high-fidelity scenarios for ADS testing has been proposed [78]. Incorporating formal verification methods, this framework ensures accurate and diverse scenario generation, improving the reliability of virtual testing. This method addresses unknown hazardous scenarios as outlined in ISO 21448.

Research focusing on generating hazardous test cases involving the ego vehicle's liability in junction scenes has been conducted [79]. A cost-based controller combined with an iterative framework is developed to produce critical scenarios involving complex traffic interactions. This approach aids in identifying potential performance insufficiencies in ADS under challenging conditions, such as intersections where liability and right-of-way rules are significant. The study is limited to junction scenarios, indicating the need for further research to generalize the methodology to other driving contexts.

A method for evaluating LiDAR performance under rainfall conditions has been proposed [80]. By analyzing risk factors affecting LiDAR, such as signal absorption and scattering by raindrops, the study highlights the impact of environmental conditions on sensor performance. This emphasizes the importance of incorporating environmental factors into safety validation, directly supporting SOTIF analysis and testing. Limitations include reliance on specific rainfall simulations, which may not capture all natural variations.

The Safety Shell, a multi-channel architecture designed to handle functional insufficiencies in automated driving, has been introduced [81]. By integrating redundant ADS channels with distinct world models and motion planning functionalities, the Safety Shell enhances safety and availability, addressing runtime uncertainties and functional insufficiencies. Challenges include computational demands and the need for optimal parameter tuning to balance safety and functionality.

The use of large language models to support hazard analysis and risk assessment (HARA) has been explored [82]. The study demonstrates that language models, when provided with structured prompts, can generate relevant safety goals, improving the efficiency of the HARA process. While this approach aligns with ISO 21448 by automating aspects of safety requirement engineering, limitations involve the risk of inaccuracies due to model limitations and the necessity for human oversight.

Gaps in ISO 21448 related to external human–machine interfaces in automated vehicles have been identified [83]. The study proposes extending the standard to include socio-technical perspectives, emphasizing problem space selection and vehicle-to-everything interaction analysis. This work highlights the importance of integrating human factors into SOTIF analysis to ensure thorough safety evaluations.

A validation testing method tailored for Highway Assistance functions has been proposed [84]. By incorporating multidimensional elements of the target market and developing second-level acceptance criteria, this method addresses unknown risks specific to highway scenarios, advancing SOTIF compliance by ensuring that validation processes are context-specific and thorough. Limitations involve reliance on target market data and balancing simulation with real-world testing due to resource constraints.

A methodology for integrating SOTIF into a validation tool suite using an ontology-driven framework for scenario description has been presented [85]. By operationalizing SOTIF principles through scenario-based validation and standardized metrics, this approach enhances the identification and management of hazardous scenarios. Challenges include scalability and dependence on existing standards, which may limit flexibility.

A method for validating perception performance insufficiencies through fault injection has been introduced [86]. By systematically injecting performance insufficiencies into the perception subsystem, the approach evaluates ADS behavior under degraded conditions. This method addresses ISO 21448 requirements by targeting unknown and hazardous scenarios resulting from perception insufficiencies. Limitations include reliance on expert judgment for calibrating factors and computational intensity of simulations.

Future research should focus on real-world validation, scalability of methodologies, and broader applicability across different operational design domains. Integrating these advancements will strengthen safety assurance and support the development of more robust and reliable ADS.

# 4. Future Research Directions

In this section, future research directions discovered during the SLR that hold significant relevance to SOTIF are briefly presented. These future research directions are derived from insights gathered from research papers encountered during the SLR, even if they were eventually excluded.

## 4.1. ML Models in Handling Situational Awareness Scenarios

Situation awareness in automated driving refers to the vehicle's ability to perceive and understand its environment in real-time. This includes recognizing elements in the surroundings, comprehending their significance, and anticipating their future states. It involves the use of sensors such as GPS, lidar, radar, and cameras to gather data about the surroundings [87].

Automated vehicles operate in an evolving environment where the ODD cannot be fully specified or anticipated. ML models, particularly DNNs, are essential for processing sensor data to identify and interpret environmental elements. Factors such as seasonal changes, sensor degradation, and variations in traffic patterns can cause distributional shifts, leading to a mismatch between training data and real-world data, thereby degrading ML model performance. DNNs are particularly brittle, with small input data perturbations potentially causing significant output deviations. This brittleness can result from changes in weather, lighting conditions, or adversarial inputs, posing safety risks by misclassifying critical elements [88].

The performance of ML models in situational awareness also depends on their ability to generalize and remain robust under varying conditions. Generalization refers to the model's capacity to handle previously unseen events accurately, while robustness indicates consistent performance despite similar input conditions [89]. However, these models face significant challenges, including model and data uncertainty due to incomplete requirement analysis, insufficient or biased training data, and real-world variations [88].

Safety concerns related to using ML models in safety-critical perception tasks include the inability of ML-based algorithms to learn semantic or causal relationships, leading to a focus on correlations in data rather than understanding the underlying reasons for predictions. This black-box behavior of ML models poses a challenge for evaluating safety aspects because it is difficult to comprehend how these models arrive at their decisions, making it hard to ensure their reliability in critical situations. Additionally, the dependence on labeled datasets for training DNNs introduces risks if the labeling quality is insufficient, potentially leading to misleading results during testing and impacting the overall performance and safety of the system [90].

ML models often lack robustness and fail to generalize across all scenarios, with standard performance metrics not adequately accounting for the severity of errors in safety-critical contexts. Safety assurance requires acknowledging these limitations, implementing both design-time and operation-time controls, and building a convincing safety case supported by iterative evidence to define and manage acceptable levels of residual risk [91].

In scenarios where ML models cannot handle specific situations in automated driving, context awareness method, which uses the Active Inference (ActInf) model, can address these weaknesses by integrating contextual data and auxiliary inputs to enhance prediction accuracy and decision-making. ActInf employs a partially observable Markov decision process (POMDP) to manage uncertainties and partial observations, allowing vehicles to infer critical information from the behaviors of other road users. This method improves safety and adaptability in real-time without extensive retraining. For instance, if a stop sign is obscured or missing, the ActInf model can infer its presence by observing cautious behaviors from other vehicles and pedestrians, ensuring accurate and safe navigation [92].

Employing supervised learning techniques maximizes black-box test coverage within a specified timeframe, improving the robustness of ML models in diverse driving scenarios. Quantum machine learning offers potential for developing ML models for safety-critical applications such as automated driving. Exploring quantum techniques could address the limitations of traditional ML models in complex driving scenarios. Refining AI hyperparameters, analyzing dataset adequacy, and systematizing AI failure modes can improve ML model performance in automated driving systems. Application-specific approaches tailored to driving scenarios are crucial for identifying and assessing corner cases. Customizing solutions for road challenges can improve ML models' ability to handle complex situations. To mitigate risks from adversarial attacks exploiting corner cases, ML models can be strengthened through adversarial training, robust optimization, and anomaly detection [93].

Moreover, fuzzy logic systems handle uncertainty and imprecision well, making them ideal for modeling human behavior and environmental factors in driving scenarios. Unlike traditional ML models, fuzzy logic systems process information more quickly, providing faster responses in real-time applications. Studies have shown that fuzzy logic can achieve comparable predictive accuracy to ML models while offering greater robustness and interpretability. Integrating fuzzy logic with ML approaches could therefore lead to reliable and efficient situational awareness systems in automated vehicles [94].

## 4.2. Implication of Uncertainty in ML-Based Perception Algorithms for SOTIF Analysis

Uncertainty is an inherent characteristic of ML algorithms, as they rely on statistical inference and probabilistic reasoning to make predictions. However, uncertainty arising from sensor limitations, algorithm complexities, and the stochastic nature of real-world scenarios can significantly affect the reliability and safety of ML-based perception systems. Without explicitly considering uncertainty in SOTIF analysis, important factors contributing to functional insufficiencies and the safety of the ADS may be overlooked.

Uncertainty manifests in diverse forms, including incomplete or noisy sensor data, ambiguities in object detection and classification, and challenges in predicting the behavior of dynamic environmental elements. These uncertainties can lead to incorrect decisions, faulty interpretations of surroundings, and inadequate responses to critical situations. Therefore, understanding and effectively managing uncertainty is essential for the safety and dependability of ML-based perception systems.

Recent works have addressed the complexities of uncertainty in ML-based perception systems within the context of SOTIF. Notably, Adee et al. [31] developed methods for modeling uncertainty, and further analyzed these methods in another study [60]. Burton and Herd [95] looked into how to ensure the safety of ML-based systems. Studies on evaluating perception for SOTIF [96] and analyzing perception architectures [97] have provided valuable insights. Efforts to measure and reduce SOTIF risks, including novel approaches to SOTIF entropy [72], managing ML uncertainties to meet safety standards [98], and digital modeling for SOTIF software interactions [99], have provided a foundation for future research.

Future research should focus on investigating and demonstrating the implications of uncertainty in ML-based perception algorithms for SOTIF analysis. Empirical research and theoretical frameworks can be employed to explore how uncertainty impacts the performance and safety of ML-based perception systems across diverse scenarios and operating conditions. Subjecting ML algorithms to different sources of uncertainty, including sensor limitations, algorithmic complexities, and environmental dynamics, insights can be gained into the functional insufficiencies that arise and their potential consequences on system safety.

## 4.3. Framework for Modeling SOTIF Scenarios and Test Case Generation

The ISO 21448 [3] standard provided guidance on the structured approach for modeling SOTIF scenarios and test case generation.

The approach involves a series of steps, starting with defining the ODD of the system and identifying triggering conditions that could impact its performance. Subsequently, SOTIF scenarios are then formulated based on these potential functional insufficiencies. Concrete test scenarios derived from these conditions enable the evaluation of the system's SOTIF performance. In this context, the triggering condition, combined with the operational situation, creates the test scenario, leading directly to unintended vehicle behavior. The pass–fail criteria are established in response to the automated vehicle's behavior in these scenarios. Figure 5 illustrates the process by which a triggering condition results in unintended behavior due to functional insufficiencies or foreseeable misuse, and how this informs the determination of pass–fail criteria based on the vehicle's response.

Test scenarios are designed to represent situations with potential functional insufficiencies, involving key system elements in perception, planning, and execution of driving tasks. Selection is based on hazardous scenarios. Elements denote the components or subsystems of the ADS involved in these tasks. Targets specify the safety goals or performance criteria for evaluating the system's ability to mitigate or avoid safety-critical hazards and risks. Action describes the control measures taken by the system to address these hazards and risks. Events identify specific scenarios or situations that may lead to safety-critical hazards and risks [3].

By deriving test cases or use cases based on different triggering conditions, the system can be evaluated for its ability to detect and respond to potential hazards. The scope of SOTIF scenarios encompasses a range of factors, including environmental factors such as weather, lighting, and road conditions, as well as driver behavior and system constraints.

Wu et al. [73] developed a hierarchy for SOTIF scenario hierarchy for risk assessment and test case generation. Ploeg et al. [100] developed a safety assessment framework that integrates FuSa design with a scenario-based approach. Piazzoni et al. [101] introduced a framework for virtual scenario-based testing, highlighting the importance of designing specific test cases to identify limitations and inconsistencies in the ADS.

Birkemeyer conducted a series of studies, as referenced in [13, 15, 102]. They presented a method for SOTIF-compliant scenario generation using semi-concrete scenarios and parameter sampling [13], identified gaps in scenario generation techniques with respect to SOTIF requirements [15], and introduced feature-interaction sampling for scenario-based testing [102].

Furthermore, Bannour et al. [103] focused on symbolic model-based design for generating logical scenarios,

**FIGURE 5** Flowchart of SOTIF scenario development and test case generation process. (Data taken from Ref. [3].)

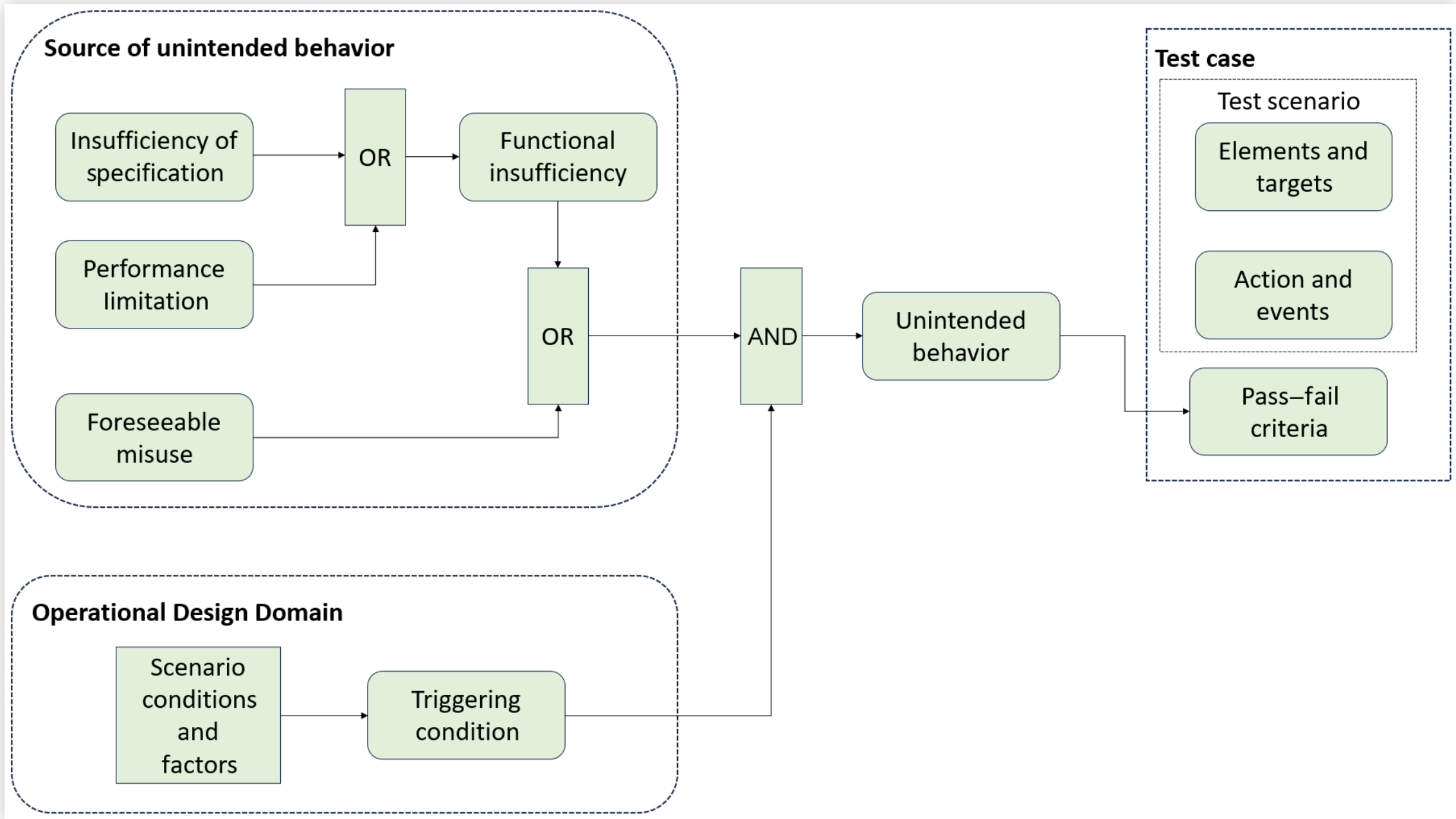

while Rajesh et al. [74] and Zhao et al. [75] explored unknown–unsafe scenario generation and constructing complex traffic scenarios, respectively, enhancing the diversity and criticality of the SOTIF scenario library.

# 5. Conclusion

This paper presents the results of a SLR on SOTIF for ADS, following the PRISMA guidelines. The review involved developing a search strategy, defining exclusion criteria, and selecting and assessing the quality of relevant studies published between 2018 and 2023, using the Parsifal tool.

Analyzing 51 research papers, this study provides an overview of the current state of research on SOTIF for ADS. The main contributions are detailed in Section 3, which includes a literature summary, acknowledged limitations, a comparative analysis of methods supporting SOTIF, and supplementary insights from recent publications addressing identified research gaps.

The literature summary highlights key thematic areas: holistic safety approaches combining FuSa and SOTIF, methods for addressing uncertainties, hazard identification and risk assessment techniques, safety requirements for ML-based perception, and scenario-based SOTIF analysis. Limitations acknowledged in the literature include challenges in validation frameworks, scope and applicability limitations of proposed approaches, data quality and simulation challenges, and human factors and ethical considerations.

The comparative analysis emphasizes the need for a multifaceted approach to address safety challenges associated with SOTIF in ADS. It evaluates the effectiveness, challenges, and practical applicability of various methods and frameworks, highlighting the importance of integrating different methodologies tailored to specific challenges. The analysis identifies gaps in practical applicability and scalability, particularly concerning data availability and computational resources.

Supplementary insights from recent publications extend the literature scope by providing findings that directly address the identified research gaps. These studies present advancements in hazard identification, scenario-based testing, handling of functional insufficiencies, integration of human factors, and validation methodologies related to SOTIF.

Critical factors for successful SOTIF implementation include risk assessment and mitigation, system design, and testing and validation. Safety challenges involve unknown scenarios, human factors, sensor limitations, and system complexity. Research gaps identified from the SLR include a limited focus on human factors and the absence of unified evaluation criteria for SOTIF measures.

Based on the findings, this paper recommends that future research focus on developing adaptable and scalable methodologies that can function with limited data and computational resources. Addressing human factors and ensuring real-world applicability are important to bridge current gaps. Establishing unified evaluation criteria for SOTIF measures would enhance consistency and comparability across studies and applications. Integrating advancements from recent studies into practical applications will strengthen safety assurance in ADS and support the development of more robust and reliable systems.

Future research directions include exploring ML models in handling situational awareness scenarios, investigating the implications of uncertainty in ML-based perception algorithms for SOTIF analysis, and developing frameworks for modeling SOTIF scenarios and generating test cases.

The limitations of this paper include the focus on literature published between 2018 and 2023, which may exclude relevant research outside this period, and the exclusion of certain publications due to the review methodology's criteria, potentially restricting the scope of findings.

In summary, implementing SOTIF measures in ADS requires an approach that integrates various methodologies tailored to specific challenges. Careful selection and combination of appropriate methods can enhance the safety and reliability of ADS, facilitating wider acceptance and deployment. This SLR provides a foundation for informed decision-making regarding suitable methodologies for SOTIF, supporting researchers and practitioners in advancing the field.

## Acknowledgements

This research was funded by the Institute for Driver Assistance and Connected Mobility (IFM) at Kempten University of Applied Sciences. The IFM specializes in the development and validation of driver assistance systems and connected mobility, focusing on FuSa, cybersecurity, and testing methodologies. I extend my gratitude to my team at IFM for their valuable support and contributions in the areas of functional safety and cybersecurity.

## Contact Information

**Milin Patel**
Institute for Advanced Driver Assistance Systems and Connected Mobility, Benningen, Bavaria, Germany
milin.patel@hs-kempten.de

**Rolf Jung**
Kempten University of Applied Science, Kempten, Bavaria, Germany
rolf.jung@hs-kempten.de

**Marzana Khatun**
University of Ulm, Ulm, Baden Wuerttemberg, Germany
marzana.khatun@alumni.uni-ulm.de

## References

1. Burton, S., Habli, I., Lawton, T., McDermid, J. et al., "Mind the Gaps: Assuring the Safety of Autonomous Systems from an Engineering, Ethical, and Legal Perspective," *Artificial Intelligence* 279 (2020): 103201, doi:https://doi.org/10.1016/j.artint.2019.103201.
2. ISO, "ISO 26262, Road Vehicles—Functional Safety," 2nd ed., 01.040.43 43.040.10, accessed September 15, 2022, https://www.iso.org/standard/68383.html.
3. ISO, "ISO 21448, Road Vehicles—Safety of the Intended Functionality," 1st ed., Switzerland, 43.040.10 Electrical and Electronic Equipment, accessed January 31, 2023, https://www.iso.org/standard/77490.html.
4. Wang, H., Shao, W., Sun, C., Yang, K. et al., "A Survey on an Emerging Safety Challenge for Autonomous Vehicles: Safety of the Intended Functionality," *Engineering* 33 (2024): 17-34, doi:https://doi.org/10.1016/j.eng.2023.10.011.
5. Takács, Á., Drexler, D.A., Galambos, P., Rudas, I.J. et al., "Assessment and Standardization of Autonomous Vehicles," in *2018 IEEE 22nd International Conference on Intelligent Engineering Systems (INES)*, Las Palmas de Gran Canaria, Spain, 2018, 185-192.
6. Birch, J. et al., "A Structured Argument for Assuring Safety of the Intended Functionality (SOTIF)," in *Computer Safety, Reliability, and Security. SAFECOMP 2020 Workshops*, Lisbon, Portugal, 2020, 408-414.
7. Zhu, R., Gu, A., Wu, Z., Liu, B. et al., "Research on SOTIF of Automatic Driving System," in *2022 14th International Conference on Measuring Technology and Mechatronics Automation (ICMTMA)*, Changsha, China, 2022, 228-231.
8. Xu, S. et al., "A Review of SOTIF Research for Human-Machine Driving Mode Switch of Intelligent Vehicles," in *2022 6th CAA International Conference on Vehicular Control and Intelligence (CVCI)*, Nanjing, China, 2022, 1-6.
9. Hoss, M., Scholtes, M., and Eckstein, L., "A Review of Testing Object-Based Environment Perception for Safe Automated Driving," *Automotive Innovation* 5, no. 3 (2022): 223-250, doi:https://doi.org/10.1007/s42154-021-00172-y.
10. Zhao, T., Yurtsever, E., Paulson, J.A., and Rizzoni, G., "Formal Certification Methods for Automated Vehicle Safety Assessment," *IEEE Transactions on Intelligent Vehicles* 8, no. 1 (2023): 232-249, doi:https://doi.org/10.1109/TIV.2022.3170517.
11. Saberi, A.K., Hegge, J., Fruehling, T., and Groote, J.F., "Beyond SOTIF: Black Swans and Formal Methods," in *2020 IEEE International Systems Conference (SysCon)*, Montreal, QC, Canada, 2020, 1-5.
12. Cao, L., He, Y., Luo, Y., and Chen, J., "Layered SOTIF Analysis and 3σ-Criterion-Based Adaptive EKF for Lidar-Based Multi-Sensor Fusion Localization System on Foggy Days," *Remote Sensing* 15, no. 12 (2023): 3047, doi:https://doi.org/10.3390/rs15123047.
13. Birkemeyer, L., Fuchs, J., Gambi, A., and Schaefer, I., "SOTIF-Compliant Scenario Generation Using Semi-Concrete Scenarios and Parameter Sampling," in *2023 IEEE 26th International Conference on Intelligent Transportation Systems (ITSC)*, Bilbao, Spain, 2023, 2139-2144.
14. Zhang, X. et al., "Finding Critical Scenarios for Automated Driving Systems: A Systematic Mapping Study," *IEEE Transactions on Software Engineering* 49, no. 03 (2023): 991-1026, doi:https://doi.org/10.1109/TSE.2022.3170122.
15. Birkemeyer, L., King, C., and Schaefer, I., "Is Scenario Generation Ready for SOTIF? A Systematic Literature Review," in *2023 IEEE 26th International Conference on Intelligent Transportation Systems (ITSC)*, Bilbao, Spain, 2023, 472-479.

16. Kitchenham, B., "Evidence-Based Software Engineering and Systematic Literature Reviews," in *Product-Focused Software Process Improvement: 7th International Conference, PROFES 2006*, Amsterdam, the Netherlands, 2006, 3.
17. Page, M.J. et al., "PRISMA 2020 Explanation and Elaboration: Updated Guidance and Exemplars for Reporting Systematic Reviews," *BMJ* 372 (2021): n160.
18. Ataei, P. and Litchfield, A., "The State of Big Data Reference Architectures: A Systematic Literature Review," *IEEE Access* 10 (2022): 113789-113807, doi:https://doi.org/10.1109/ACCESS.2022.3217557.
19. Sousa, R.B., Sobreira, H.M., and Moreira, A.P., "A Systematic Literature Review on Long-Term Localization and Mapping for Mobile Robots," *Journal of Field Robotics* 40, no. 5 (2023): 1245-1322, doi:https://doi.org/10.1002/rob.22170.
20. Zhang, J. and Li, J., "Testing and Verification of Neural-Network-Based Safety-Critical Control Software: A Systematic Literature Review," *Information and Software Technology* 123 (2020): 106296, doi:https://doi.org/10.1016/j.infsof.2020.106296.
21. Pérez-Verdejo, M.J., Sánchez-García, A.J., and Ocharán-Hernández, J.O., "A Systematic Literature Review on Machine Learning for Automated Requirements Classification," in *2020 8th International Conference in Software Engineering Research and Innovation (CONISOFT)*, Chetumal, Mexico, 2020, 21-28.
22. Freitas, V., "ParsifalPerform Systematic Literature Reviews: Perform Systematic Literature Reviews," 2021, accessed July 31, 2023, https://parsif.al/blog/.
23. José, L.O., "Chapter 7—Other Academic Search Engines," in Ortega, J.L. (ed.), *Chandos Information Professional Series, Academic Search Engines* (Oxford: Chandos Publishing, 2014), 143-157.
24. Kirovskii, O.M. and Gorelov, V.A., "Driver Assistance Systems: Analysis, Tests and the Safety Case. ISO 26262 and ISO PAS 21448," *IOP Conference Series: Materials Science and Engineering* 534 (2019): 12019.
25. Collin, A., Bilka, A., Pendleton, S., and Tebbens, R.D., "Safety of the Intended Driving Behavior Using Rulebooks," in *2020 IEEE Intelligent Vehicles Symposium (IV)*, Las Vegas, NV, 2020, 136-143.
26. Kinalzyk, D., "SOTIF Process and Methods in Combination with Functional Safety," in Yilmaz, M., Clarke, P., Messnarz, R., and Reiner, M. (eds.), *Systems, Software and Services Process Improvement* (Cham, Switzerland: Springer International Publishing, 2021), 612-623.
27. Poddey, A., Brade, T., Stellet, J.E., and Branz, W., "On the Validation of Complex Systems Operating in Open Contexts," arXiv preprint arXiv:1902.10517, 2019, https://api.semanticscholar.org/CorpusID:67855393.
28. Drage, T., Lim, K.L., Hai Koh, J.E., Gregory, D. et al., "Integrated Modular Safety System Design for Intelligent Autonomous Vehicles," in *2021 IEEE Intelligent Vehicles Symposium (IV)*, Nagoya, Japan, 2021, 258-265.
29. Feth, P. et al., "Multi-Aspect Safety Engineering for Highly Automated Driving," in *Computer Safety, Reliability, and Security: 37th International Conference, SAFECOMP 2018*, Västerås, Sweden, 2018, 59-72.
30. Lotto, K., Nagler, T., and Radic, M., "Modeling Stochastic Data Using Copulas for Applications in the Validation of Autonomous Driving," *Electronics* 11, no. 24 (2022): 4154, doi:https://doi.org/10.3390/electronics11244154.
31. Adee, A., Munk, P., Gansch, R., and Liggesmeyer, P., "Uncertainty Representation with Extended Evidential Networks for Modeling Safety of the Intended Functionality (SOTIF)," in *European Safety and Reliability Conference (ESREL2020)*, Venice, Italy, 2020, 4148-4156.
32. Peng, L., Wang, H., and Li, J., "Uncertainty Evaluation of Object Detection Algorithms for Autonomous Vehicles," *Automotive Innovation* 4, no. 3 (2021): 241-252, doi:https://doi.org/10.1007/s42154-021-00154-0.
33. Henriksson, J. et al., "Out-of-Distribution Detection as Support for Autonomous Driving Safety Lifecycle," in Ferrari, A. and Penzenstadler, B. (eds.), *Requirements Engineering: Foundation for Software Quality* (Cham, Switzerland: Springer International Publishing, 2023), 233-242.
34. Babisch, S., Neurohr, C., Westhofen, L., Schoenawa, S. et al., "Leveraging the GIDAS Database for the Criticality Analysis of Automated Driving Systems," *Journal of Advanced Transportation* 2023 (2023): 1349269, doi:https://doi.org/10.1155/2023/1349269.
35. Cieslik, I., Expósito Jiménez, V.J., Martin, H., Scharke, H. et al., "State of the Art Study of the Safety Argumentation Frameworks for Automated Driving System," in *Computer Safety, Reliability, and Security. SAFECOMP 2022 Workshops*, Munich, Germany, 2022, 178-191.
36. Zhang, S., Tang, T., and Liu, J., "A Hazard Analysis Approach for the SOTIF in Intelligent Railway Driving Assistance Systems Using STPA and Complex Network," *Applied Sciences* 11, no. 16 (2021): 7714, doi:https://doi.org/10.3390/app11167714.
37. Delaus, M., Taft, G.D.H., Dhavaleswarapu, D., Hao, C. et al., "Bayesian Network Analysis of Safety of Intended Functionality of System Designs," 2022/0067550 A1, United States 17/011,384, March 2022.
38. Qin, K., Wang, Y., Liu, L., Xia, X. et al., "Analysis and Research on SOTIF of Typical L3 Autopilot System," in *2022 7th International Conference on Cyber Security and Information Engineering (ICCSIE)*, Brisbane, Australia, 2022, 82-86.
39. Wang, B., Luo, Y., Zhong, Z., and Li, K., "Robust Non-Fragile Fault Tolerant Control for Ensuring the Safety of the Intended Functionality of Cooperative Adaptive Cruise Control," *IEEE Transactions on Intelligent Transportation Systems* 23, no. 10 (2022): 18746-18760, doi:https://doi.org/10.1109/TITS.2022.3161012.
40. Yu, W., Li, J., Peng, L.-M., Xiong, X. et al., "SOTIF Risk Mitigation Based on Unified ODD Monitoring for Autonomous Vehicles," *Journal of Intelligent and Connected Vehicles* 5, no. 3 (2022): 157-166, doi:https://doi.org/10.1108/JICV-04-2022-0015.
41. Kramer, B., Neurohr, C., Büker, M., Böde, E. et al., "Identification and Quantification of Hazardous Scenarios for Automated Driving," in Zeller, M., and Höfig, K. (eds.),

*Model-Based Safety and Assessment* (Cham, Switzerland: Springer International Publishing, 2020), 163-178.

42. Warg, F. et al., "The Quantitative Risk Norm - A Proposed Tailoring of HARA for ADS," in *2020 50th Annual IEEE/IFIP International Conference on Dependable Systems and Networks Workshops (DSN-W)*, Valencia, Spain, 2020, 86-93.
43. Junfeng, L., Yunshuang, Z., Shuai, Z., Chao, C. et al., "A Research on SOTIF of LKA Based on STPA," in *2022 IEEE International Conference on Real-Time Computing and Robotics (RCAR)*, Guiyang, China, 2022, 396-400.
44. Celik, E.A., Cârlan, C., Abdulkhaleq, A., Bauer, F. et al., "Application of STPA for the Elicitation of Safety Requirements for a Machine Learning-Based Perception Component in Automotive," in Trapp, M., Saglietti, F., Spisländer, M., and Bitsch, F. (eds.), *Computer Safety, Reliability, and Security* (Cham, Switzerland: Springer International Publishing, 2022), 319-332.
45. Borg, M. et al., "Ergo, SMIRK Is Safe: A Safety Case for a Machine Learning Component in a Pedestrian Automatic Emergency Brake System," *Software Quality Journal* 31, no. 2 (2023): 335-403, doi:https://doi.org/10.1007/s11219-022-09613-1.
46. Kaneko, T., Takahashi, Y., Yamaguchi, S., Hashimoto, J. et al., "Safety and Risk Analysis and Evaluation Methods for DNN Systems in Automated Driving," in Virvou, M., Saruwatari, T., and Jain, L.C. (eds), *Knowledge-Based Software Engineering: 2022* (Cham, Switzerland: Springer International Publishing, 2023), 83-96.
47. Hacker, L. and Seewig, J., "Insufficiency-Driven DNN Error Detection in the Context of SOTIF on Traffic Sign Recognition Use Case," *IEEE Open Journal of Intelligent Transportation Systems* 4 (2023): 58-70, doi:https://doi.org/10.1109/OJITS.2023.3236531.
48. Shao, W., Li, J., and Wang, H., "Self-Aware Trajectory Prediction for Safe Autonomous Driving," in *2023 IEEE Intelligent Vehicles Symposium (IV)*, Anchorage, AK, 2023.
49. Jhung, J., Suk, H., Park, H., and Kim, S., "Hardware Accelerators for Autonomous Vehicles," in Mishra, A., Cha, J., Park, H., and Kim, S. (eds.), *Artificial Intelligence and Hardware Accelerators* (Cham, Switzerland: Springer International Publishing), 2023, 269-317.
50. Peng, L., Li, J., Shao, W., and Wang, H., "PeSOTIF: A Challenging Visual Dataset for Perception SOTIF Problems in Long-Tail Traffic Scenarios," in *2023 IEEE Intelligent Vehicles Symposium (IV)*, Anchorage, AK, 2023, 1-8.
51. Maier, R., Grabinger, L., Urlhart, J., and Mottok, J., "Towards Causal Model-Based Engineering in Automotive System Safety," in *International Symposium on Model-Based Safety and Assessment*, Munich, Germany, 2022, 116-129.
52. Meyer, M.-A. et al., "Scenario-and Model-Based Systems Engineering Procedure for the SOTIF-Compliant Design of Automated Driving Functions," in *2022 IEEE Intelligent Vehicles Symposium (IV)*, Aachen, Germany, 2022, 1599-1604.
53. Khatun, M., Caldeira, G.B., Jung, R., and Glaß, M., "An Optimization and Validation Method to Detect the Collision Scenarios and Identifying the Safety Specification of Highly Automated Driving Vehicle," in *2021 21st International Conference on Control, Automation and Systems (ICCAS)*, Jeju, Korea, 2021, 1570-1575.
54. Wotawa, F., Peischl, B., Klück, F., and Nica, M., "Quality Assurance Methodologies for Automated Driving," *Elektrotechnik und Informationstechnik* 135, no. 4 (2018): 322-327, doi:https://doi.org/10.1007/s00502-018-0630-7.
55. Schnellbach, A. and Griessnig, G., "Development of the ISO 21448," in Walker, A., O'Connor, R., and Messnarz, R. (eds.), *Systems, Software and Services Process Improvement* (Cham, Switzerland: Springer International Publishing, 2019), 585-593.
56. Roshdi, M., Nayeer, N., Elmahgiubi, M., Agrawal, A. et al., "A Unified Evaluation Framework for Autonomous Driving Vehicles," in *2020 IEEE Intelligent Vehicles Symposium (IV)*, Las Vegas, NV, 2020, 1277-1282.
57. Kaiser, B., Dion, B., Tolchinsky, I., Le Sergent, T. et al., "An AEBS Use Case for Model-Based System Design Integrating Safety Analyses and Simulation," in Seguin, C., Zeller, M., and Prosvirnova, T. (eds.), *Model-Based Safety and Assessment* (Cham, Switzerland: Springer, 2022), 3-20.
58. Putze, L., Westhofen, L., Koopmann, T., Böde, E. et al., "On Quantification for SOTIF Validation of Automated Driving Systems," in *2023 IEEE Intelligent Vehicles Symposium (IV)*, Anchorage, AK, 2023, 1-8.
59. Wäschle, M., Thaler, F., Berres, A., Pölzlbauer, F. et al., "A Review on AI Safety in Highly Automated Driving," *Frontiers in Artificial Intelligence* 5 (2022): 952773, doi:https://doi.org/10.3389/frai.2022.952773.
60. Adee, A., Gansch, R., and Liggesmeyer, P., "Systematic Modeling Approach for Environmental Perception Limitations in Automated Driving," in *2021 17th European Dependable Computing Conference (EDCC)*, Munich, Germany, 2021, 103-110.
61. Duan, S., Li, W., Chen, J., Li, Q. et al., "Influence of Fog Weather on Automotive Vision Target Detection," in *2022 6th CAA International Conference on Vehicular Control and Intelligence (CVCI)*, Nanjing, China, 2022, 1-5.
62. Chia, W.M.D., Keoh, S.L., Goh, C., and Johnson, C., "Risk Assessment Methodologies for Autonomous Driving: A Survey," *IEEE Transactions on Intelligent Transportation Systems* 23, no. 10 (2022): 16923-16939, doi:https://doi.org/10.1109/TITS.2022.3163747.
63. Duthon, P., Edelstein, N., Zelentzer, E., and Bernardin, F., "Quadsight® Vision System in Adverse Weather Maximizing the Benefits of Visible and Thermal Cameras," in *2022 12th International Conference on Pattern Recognition Systems (ICPRS)*, Saint-Etienne, France, 2022, 1-6.
64. Xie, G., Li, Y., Han, Y., Xie, Y. et al., "Recent Advances and Future Trends for Automotive Functional Safety Design Methodologies," *IEEE Transactions on Industrial Informatics* 16, no. 9 (2020): 5629-5642, doi:https://doi.org/10.1109/TII.2020.2978889.
65. Zhu, Z., Philipp, R., Hungar, C., and Howar, F., "Systematization and Identification of Triggering Conditions: A Preliminary Step for Efficient Testing of Autonomous Vehicles," in *2022 IEEE Intelligent Vehicles Symposium (IV)*, Aachen, Germany, 2022, 798-805.

66. Nouri, A., Berger, C., and Törner, F., "An Industrial Experience Report about Challenges from Continuous Monitoring, Improvement, and Deployment for Autonomous Driving Features," in *2022 48th Euromicro Conference on Software Engineering and Advanced Applications (SEAA)*, Gran Canaria, Spain, 2022, 358-365.
67. Expósito Jiménez, V.J., Martin, H., Schwarzl, C., Macher, G. et al., "Triggering Conditions Analysis and Use Case for Validation of ADAS/ADS Functions," in *Computer Safety, Reliability, and Security. SAFECOMP 2022 Workshops*, Munich, Germany, 2022, 11-22.
68. Yan, M., Chen, W., Wang, Q., Zhao, L. et al., "Human-Machine Cooperative Control of Intelligent Vehicles for Lane Keeping—Considering Safety of the Intended Functionality," *Actuators* 10, no. 9 (2021): 210, doi:https://doi.org/10.3390/act10090210.
69. Dasa, K. and Barbora, B., "Rethinking Certification for Higher Trust and Ethical Safeguarding of Autonomous Systems," in *Proceedings of the 18th International Conference on Evaluation of Novel Approaches to Software Engineering - ENASE*, Prague, Czech Republic, 2023, 158-169, doi:https://doi.org/10.5220/0011971500003464.
70. Hou, Z., Liu, H., and Zhang, Y., "Zero-day Vulnerability Inspired Hazard Assessment for Autonomous Driving Vehicles," in *2019 IEEE 19th International Conference on Communication Technology (ICCT)*, Xi'an, China, 2019, 1348-1354.
71. Qiu, M., Antesberger, T., Bock, F., and German, R., "Exploring the Impact of Scenario and Distance Information on the Reliability Assessment of Multi-Sensor Systems," in *2022 48th Euromicro Conference on Software Engineering and Advanced Applications (SEAA)*, Gran Canaria, Spain, 2022, 322-329.
72. Peng, L., Li, B., Yu, W., Yang, K. et al., "SOTIF Entropy: Online SOTIF Risk Quantification and Mitigation for Autonomous Driving," *IEEE Transactions on Intelligent Transportation Systems* 25, no. 2 (2024): 1530-1546, doi:https://doi.org/10.1109/TITS.2023.3322166.
73. Wu, S., Wang, H., Yu, W., Yang, K. et al., "A New SOTIF Scenario Hierarchy and Its Critical Test Case Generation Based on Potential Risk Assessment," in *2021 IEEE 1st International Conference on Digital Twins and Parallel Intelligence (DTPI)*, Beijing, China, 2021, 399-409.
74. Rajesh, N., van Hassel, E., and Alirezaei, M., "Unknown-Unsafe Scenario Generation for Verification and Validation of Automated Vehicles," in *2023 IEEE 26th International Conference on Intelligent Transportation Systems (ITSC)*, Bilbao, Spain, 2023, 1759-1764.
75. Zhao, S. et al., "Genetic Algorithm-Based SOTIF Scenario Construction for Complex Traffic Flow," *Automotive Innovation* 6, no. 4 (2023): 531-546, doi:https://doi.org/10.1007/s42154-023-00251-2.
76. Hou, Z., Liu, D., Yang, Y., and Liu, H., "Assessing Unknown Hazards for SOTIF Based on Twin Scenarios Empowered Autonomous Driving," *IEEE Internet of Things Journal* 11, no. 20 (2024): 32631-32644.
77. Li, Y., Wu, S., and Wang, H., "Adaptive Mining of Failure Scenarios for Autonomous Driving Systems Based on Multi-Population Genetic Algorithm," in *2024 IEEE Intelligent Vehicles Symposium (IV)*, Jeju Island, Republic of Korea, 2024, 2458-2464.
78. Hou, Z., Wang, S., Liu, H., Yang, Y. et al., "Twin Scenarios Establishment for Autonomous Vehicle Digital Twin Empowered Sotif Assessment," *IEEE Transactions on Intelligent Vehicles* 9, no. 1 (2023): 1965-1976.
79. Xie, Y., Zhang, Y., Yin, C., and Dai, K., "Generation of Ego-Liable Hazardous-Test-Cases for Validating Automated Driving Systems in Junction-Scenes," in *2024 IEEE Intelligent Vehicles Symposium (IV)*, Jeju Island, Republic of Korea, 2024, 1157-1164.
80. Zhang, X., Shen, Y., and Shen, L., "Performance Evaluation Method of LiDAR in Rainfall Conditions," in *2024 Photonics & Electromagnetics Research Symposium (PIERS)*, Chengdu, China, 2024, 1-8.
81. Hanselaar, C.A.J., Silvas, E., Terechko, A., and Heemels, W., "The Safety Shell: An Architecture to Handle Functional Insufficiencies in Automated Driving," *IEEE Transactions on Intelligent Transportation Systems* 25, no. 7 (2024): 7522-7540.
82. Nouri, A., Cabrero-Daniel, B., Törner, F., Sivencrona, H. et al., "Engineering Safety Requirements for Autonomous Driving with Large Language Models," in *2024 IEEE 32nd International Requirements Engineering Conference (RE)*, Reykjavik, Iceland, 2024, 218-228.
83. Okada, M. and Gallina, B., "Safety of the Intended Functionality of External Human Interfaces: Gaps and Research Agenda," in *2024 IEEE 48th Annual Computers, Software, and Applications Conference (COMPSAC)*, Osaka, Japan, 2024, 578-583.
84. Guo, K., Zhang, J., Shi, J., Zhang, Z. et al., "Research and Practice on the Validation Testing Method of Safety of the Intended Functionality for High Way Assist Function," in *International Conference on Advances in Construction Machinery and Vehicle Engineering*, Shanghai, China, 2023, 1025-1034.
85. Expósito Jiménez, V.J. et al., "Safety of the Intended Functionality Concept Integration into a Validation Tool Suite," *ACM SIGAda Ada Letters* 43, no. 2 (2024): 69-72.
86. Expósito Jiménez, V.J., Macher, G., Watzenig, D., and Brenner, E., "Safety of the Intended Functionality Validation for Automated Driving Systems by Using Perception Performance Insufficiencies Injection," *Vehicles* 6, no. 3 (2024): 1164-1184.
87. Ignatious, H.A., El-Sayed, H., Khan, M.A., and Mokhtar, B.M., "Analyzing Factors Influencing Situation Awareness in Autonomous Vehicles—A Survey," *Sensors* 23, no. 8 (2023): 4075.
88. Abrecht, S., Hirsch, A., Raafatnia, S., and Woehrle, M., "Deep Learning Safety Concerns in Automated Driving Perception," arXiv preprint arXiv:2309.03774, 2023.
89. Wang, J., Li, H., Wang, H., Pan, S.J. et al., "Trustworthy Machine Learning: Robustness, Generalization, and Interpretability," in *Proceedings of the 29th ACM SIGKDD Conference on Knowledge Discovery and Data Mining*, Long Beach, CA, 2023, 5827-5828.

90. Willers, O., Sudholt, S., Raafatnia, S., and Abrecht, S., "Safety Concerns and Mitigation Approaches Regarding the Use of Deep Learning in Safety-Critical Perception Tasks," in *Computer Safety, Reliability, and Security. SAFECOMP 2020 Workshops: DECSoS 2020, DepDevOps 2020, USDAI 2020, and WAISE 2020*, Lisbon, Portugal, 2020, 336-350.
91. Burton, S., Hellert, C., Hüger, F., Mock, M. et al., "Safety Assurance of Machine Learning for Perception Functions," in Fingscheidt, T., Gottschalk, H., and Houben, S. (eds.), *Deep Neural Networks and Data for Automated Driving: Robustness, Uncertainty Quantification, and Insights Towards Safety* (Cham, Switzerland: Springer International Publishing, 2022), 335-358.
92. Zaboli, A., Hong, J., Kwon, J., and Moore, J., "A Survey on Cyber-Physical Security of Autonomous Vehicles Using a Context Awareness Method," *IEEE Access* 11 (2023): 136706-136725, doi:https://doi.org/10.1109/ACCESS.2023.3338156.
93. Neto, A.V.S., Camargo, J.B., Almeida, J.R., and Cugnasca, P.S., "Safety Assurance of Artificial Intelligence-Based Systems: A Systematic Literature Review on the State of the Art and Guidelines for Future Work," *IEEE Access* 10 (2022): 130733-130770.
94. Ferenc, G., Timotijević, D., Tanasijević, I., and Simić, D., "Towards Enhanced Autonomous Driving Takeovers: Fuzzy Logic Perspective for Predicting Situational Awareness," *Applied Sciences* 14, no. 13 (2024): 5697, doi:https://doi.org/10.3390/app14135697.
95. Burton, S. and Herd, B., "Addressing Uncertainty in the Safety Assurance of Machine-Learning," *Frontiers in Computer Science* 5 (2023): 1132580.
96. Chu, J., Zhao, T., Jiao, J., Yuan, Y. et al., "SOTIF-Oriented Perception Evaluation Method for Forward Obstacle Detection of Autonomous Vehicles," *IEEE Systems Journal* 17, no. 2 (2023): 2319-2330, doi:https://doi.org/10.1109/JSYST.2023.3234200.
97. Kurzidem, I., Saad, A., and Schleiss, P., "A Systematic Approach to Analyzing Perception Architectures in Autonomous Vehicles," in *Model-Based Safety and Assessment: 7th International Symposium, IMBSA 2020*, Lisbon, Portugal, 2020, 149-162.
98. Vasudevan, V., Abdullatif, A., Kabir, S., and Campean, F., "A Framework to Handle Uncertainties of Machine Learning Models in Compliance with ISO 26262," in Jansen, T., Jensen, R., Mac Parthaláin, N., and Lin, C.M. (eds.), *Advances in Computational Intelligence Systems* (Cham, Switzerland: Springer, 2022), 508-518.
99. Wu, Z., Yang, X., Chen, P., Qu, Z. et al., "Multi-Scale Software Network Model for Software Safety of the Intended Functionality," in *2021 IEEE International Symposium on Software Reliability Engineering Workshops (ISSREW)*, Wuhan, China, 2021, 250-255.
100. Ploeg, J., de Gelder, E., Slavík, M., Querner, E. et al., "Scenario-Based Safety Assessment Framework for Automated Vehicles," arXiv preprint arXiv:2112.09366, 2021, https://api.semanticscholar.org/CorpusID:208979008.
101. Piazzoni, A., Cherian, J., Azhar, M., Yap, J. et al., "ViSTA: A Framework for Virtual Scenario-based Testing of Autonomous Vehicles," in *2021 IEEE International Conference on Artificial Intelligence Testing (AITest)*, Oxford, UK, 2021, 143-150, https://doi.org/10.1109/AITEST52744.2021.00035.
102. Birkemeyer, L., Pett, T., Vogelsang, A., Seidl, C. et al., "Feature-Interaction Sampling for Scenario-Based Testing of Advanced Driver Assistance Systems," in *Proceedings of the 16th International Working Conference on Variability Modelling of Software-Intensive Systems*, Florence, Italy, 2022.
103. Bannour, B., Niol, J., and Crisafulli, P., "Symbolic Model-Based Design and Generation of Logical Scenarios for Autonomous Vehicles Validation," in *2021 IEEE Intelligent Vehicles Symposium (IV)*, Nagoya, Japan, 2021, 215-222.

# Appendix A: Supplementary Data

All the supported data used for this SLR is available at the following link: https://dx.doi.org/10.21227/gczg-nc15.